\documentclass[preprint]{elsarticle} 

\makeatletter
	\def\ps@pprintTitle{%
 	\let\@oddhead\@empty
	\let\@evenhead\@empty
	\def\@oddfoot{\centerline{\thepage}}%
	\let\@evenfoot\@oddfoot}
\makeatother

\usepackage{etoolbox}
\patchcmd{\MaketitleBox}{\footnotesize\itshape\elsaddress\par\vskip36pt}{\footnotesize\itshape\elsaddress\par\parbox[b][36pt]{\linewidth}{\vfill\hfill\textnormal{\today}\hfill\null\vfill}}{}{}%
\patchcmd{\pprintMaketitle}{\footnotesize\itshape\elsaddress\par\vskip36pt}{\footnotesize\itshape\elsaddress\par\parbox[b][36pt]{\linewidth}{\vfill\hfill\textnormal{\today}\hfill\null\vfill}}{}{}%

\usepackage[colorlinks=true,%
            breaklinks=true,%
            linkcolor=blue,
            urlcolor=blue,%
            citecolor=blue,%
            pdftitle={Title of paper}, 
            pdfkeywords={Keywords},	
            pdfauthor={Authors},		
            bookmarksopen=false,
            pdfpagemode=UseNone]{hyperref}

\usepackage[margin=0.8in]{geometry}
\usepackage{makecell}
\usepackage[T1]{fontenc}
\usepackage[english]{babel}
\usepackage{icomma}
\usepackage[latin1]{inputenc} 
\usepackage{subfiles}
\usepackage{xcolor}
\usepackage{tabularray}
\biboptions{numbers,sort&compress}
\usepackage[justification=centering]{subcaption}

\usepackage[makeroom]{cancel}
\usepackage{amsmath}
\usepackage{pifont}

\usepackage{mathtools}
\usepackage{siunitx}
\usepackage{bbm}
\usepackage{amsmath}
\usepackage{amsfonts}
\usepackage{amsthm}
\usepackage{amssymb}
\usepackage{array}
\usepackage{algorithm} 
\usepackage{algorithmicx}
\usepackage{algpseudocode}
\usepackage{siunitx}
\usepackage{amsmath}
\usepackage{mathrsfs}
\usepackage{xpatch}

\usepackage{color}
\usepackage{url}
\usepackage{cleveref}
\usepackage{graphicx}
\usepackage{breakcites}
\usepackage{soul} 
\usepackage{enumitem}
\usepackage[title,titletoc,toc]{appendix}

\newtheoremstyle{mytheoremstyle}{5pt}{5pt}{\itshape}{}{\bfseries}{.}{.5em}{} 
\theoremstyle{mytheoremstyle}

\newtheoremstyle{myremarkstyle}{3pt}{3pt}{\itshape}{}{\bfseries}{.}{.5em}{} 
\theoremstyle{myremarkstyle}

\usepackage{pifont}

\newcommand{\Hquad}{\hspace{0.5em}}

\begin{document}
\begin{frontmatter}
    \title{Mixed Hermite--Legendre spectral method for kinetic plasma simulations}
    \author[1,2]{Opal Issan\corref{cor1}}\ead{oissan@ucsd.edu}
    \author[2]{Gian Luca Delzanno}
    \author[2]{Vadim Roytershteyn}
    
    \cortext[cor1]{Corresponding author}
    \address[1]{Department of Mechanical and Aerospace Engineering, University of California San Diego, La Jolla, CA, USA}
    \address[2]{T--5 Applied Mathematics and Plasma Physics Group, Los Alamos National Laboratory, Los Alamos, NM, USA}
    \begin{abstract}
    Kinetic collisionless plasma equations are commonly solved via spectral methods in velocity space. 
    The most commonly used spectral method is based on Hermite polynomials with a Maxwellian weight, as this basis efficiently represents near-Maxwellian distributions with relatively few degrees of freedom.
    An alternative approach uses Legendre polynomials, which are better suited for resolving strongly non-Maxwellian features.
    In this paper, we propose a mixed method that combines the Hermite and Legendre expansions. 
    The mixed method is particularly advantageous for problems in which non-Maxwellian features are localized in velocity space, such as beams and plateaus. 
    We demonstrate analytically and numerically that the mixed method conserves total mass, momentum, and energy by imposing certain constraints. 
    The numerical results show that, for the same number of degrees of freedom, the proposed mixed method can achieve improved accuracy in comparison to the individual Hermite or Legendre methods, while maintaining comparable computational cost.
    \end{abstract}	
\end{frontmatter}

\section{Introduction}\label{sec:introduction}
The numerical discretization of the Vlasov equation is a very active area of research, due to its many challenges involving significant scale separation, nonlinearities, six-dimensional phase space, and rich geometric structure. 
There are broadly three main classes of discretization methods for solving the Vlasov equation.
The most commonly used method is Particle-In-Cell (PIC)~\cite{birdsall_1991_PIC, hockney_1988_PIC}, in which the particle distribution function is sampled using a discrete set of macro-particles and the electromagnetic fields are discretized on a grid.
The main advantages of PIC are its robustness, relative simplicity, and efficient parallelization; however, it is a low-order method as it suffers from the well-known statistical noise associated with Monte Carlo methods.
The second class of methods solves the Vlasov equation using a computational grid in phase space based on finite difference~\cite{shiroto_2019_fd, carrie_2022_fd}, finite volume~\cite{filbet_2001_fv, vogman_2018_fv}, finite element~\cite{kormann_2025_fe, zaki_1988_fe}, and discontinuous Galerkin~\cite{heath_2012_dg, cheng_2014_dg, juno_2018_dg} methods. 
Lastly, the third class is spectral methods (also called \textit{transform methods}), where the velocity space is projected onto an orthogonal basis, e.g., Hermite~\cite{grad_1949_hermite}, Legendre~\cite{manzini_2016_jcp}, Fourier~\cite{klimas_1983_fourier}, Chebyshev~\cite{shoucri_1974_chebyshev}, and Laguerre~\cite{juno_2025_laguerre} functions.
Grid and spectral methods do not suffer from statistical noise (unlike PIC) and can therefore resolve fine-scale structures more easily; however, their main limitation is their high memory and computational cost due to the six-dimensional phase space, also known as the \textit{curse of dimensionality}.

Hermite-based spectral methods provide a natural framework for expanding the velocity dependence of the particle distribution function, which has led to a substantial body of literature~\cite{grad_1949_hermite, armstrong_1967_pof, grant_1967_pof, schumer_1998_jcp, delzanno_2015_jcp, camporeale_2016_cpc, koshkarov_2021_cpc, issan_2024_jcp, parker_2014_jpp, barbour_2025_jpp, filbet_2022_camc, pezzi_2019_ppcf, joyce_1971_jcp, issan_2024_pop, issan_2025_sw_closure, chapurin_2024_pop, kormann_2021_generalized_hermite}.
This is largely due to their ability to efficiently resolve near-Maxwellian features using only a small number of basis functions.
The Hermite basis functions are constructed by multiplying Hermite polynomials by a Maxwellian weight, such that a Maxwellian distribution is exactly represented by the zeroth-order Hermite basis function.
There are two types of Hermite bases in the literature, which differ by their Maxwellian weight: \textit{asymmetrically weighted}~(AW) and \textit{symmetrically weighted}~(SW). 
The AW Hermite basis has the ``fluid-kinetic'' coupling property, whereby the first three Hermite coefficients describe the fluid moments, i.e., mass, momentum, and kinetic energy. 
This can be interpreted as a more general, fluid-based hierarchy that, in the limit of an infinite number of spectral bases, converges to the kinetic description.
The AW Hermite expansion inherently conserves mass, momentum, and energy regardless of the closure term, although $L^{2}$-stability is not guaranteed~\cite{schumer_1998_jcp, camporeale_2016_cpc, delzanno_2015_jcp} and can be enforced by adding artificial collisions~\cite{funaro_2021_hermite_stability}. 
In contrast, the SW Hermite basis cannot simultaneously conserve mass, momentum, and energy regardless of the closure term, but is $L^{2}$-stable with \textit{closure by truncation} (where the last Hermite coefficient is set to zero)~\cite{schumer_1998_jcp, issan_2025_sw_closure}.
However, because both formulations rely on a Maxwellian weight, Hermite-based discretizations tend to converge slowly in the presence of strongly non-Maxwellian features, often requiring non-trivial optimization of spectral scaling and shifting parameters~\cite{schumer_1998_jcp, fatone_2022_hermite, pagliantini_2023_jcp, shao_2025_adaptive}.

The Legendre basis, on the other hand, is particularly well-suited for representing strongly non-Maxwellian distributions, where Hermite expansions converge slowly.
It consists of pure polynomials with a unit weight function, forming an orthogonal basis in velocity space.
The Legendre discretization of the Vlasov equation was first introduced by~\citet{manzini_2016_jcp}.
Similar to the AW Hermite expansion, the Legendre expansion exhibits the ``fluid-kinetic'' coupling property, and conserves mass, momentum, and energy. 
Furthermore, the Legendre basis is defined on a bounded domain, such that decreasing the domain improves the spectral accuracy for a fixed number of basis functions. 
Lastly, $L^{2}$-stability is ensured by introducing a penalty term based on the velocity Dirichlet boundary conditions of the particle distribution function. 

In this paper, we introduce a mixed method that combines the AW Hermite and Legendre spectral expansions. 
In weakly collisional and collisionless plasmas, the distribution function often develops localized non-Maxwellian features in velocity space, such as plateaus and beam-like structures.
In such scenarios, the proposed mixed method is particularly advantageous as it can efficiently capture the bulk of the distribution using the AW Hermite basis with a few basis functions, while strongly non-Maxwellian features are confined to a limited region, where the Legendre expansion is applied. 
The proposed mixed method aligns most closely with the recent paper by~\citet{chapurin_2024_pop}, which developed a hybrid method combining the AW Hermite and PIC approaches.
In the hybrid formulation of \citet{chapurin_2024_pop}, the spectral and particle components interact exclusively via the fields. 
This decoupling renders the method effectively static, potentially requiring an excessive number of degrees of freedom~(DOFs) if the initial phase-space decomposition is suboptimal. 
By contrast, the proposed mixed method dynamically redistributes information between the near-Maxwellian and non-Maxwellian components.

The main contribution of this paper is the formulation of the mixed Hermite--Legendre method for the 1D--1V Vlasov--Poisson equations. 
We show that the mixed method inherits the ``fluid-kinetic'' coupling property of both the AW Hermite and Legendre expansions, as well as their conservation of total mass, momentum, and energy. 
Conservation is ensured by restricting the coupling between the last Hermite coefficient and the Legendre coefficients to those higher than the third Legendre coefficient.
We verify numerically the analytically derived conservation properties on the following benchmark problems: linear advection, two-stream instability, and bump-on-tail instability. 
The numerical results demonstrate that the mixed method is more accurate than the AW Hermite and Legendre methods alone when non-Maxwellian features are localized in velocity space.

This paper is organized as follows. Section~\ref{sec:method_formulation} describes the 1D--1V Vlasov-Poisson equations and their discretization in velocity via the mixed method. 
In section~\ref{sec:conservation_laws}, we derive conservation laws for total mass, momentum, and energy.  
Section~\ref{sec:numerical_results} presents the mixed method numerical results tested on standard electrostatic benchmark problems, and we conclude the paper in section~\ref{sec:conclusions}.

\section{Formulation of the mixed Hermite--Legendre method}\label{sec:method_formulation}
We present the 1D--1V Vlasov--Poisson equations in section~\ref{sec:vlasov_poisson_equations}. 
The AW Hermite and Legendre bases are defined in sections~\ref{sec:hermite} and~\ref{sec:legendre}, respectively.
Section~\ref{sec:mixed_method_derived} derives the mixed method equations after discretization in velocity and strategically decomposing the particle distribution function. 
Lastly, we describe the artificial collisional operator used to improve numerical stability in section~\ref{sec:collisions}.

\subsection{Vlasov--Poisson equations}\label{sec:vlasov_poisson_equations}
Consider the 1D--1V Vlasov--Poisson equations, which model the interaction of collisionless charged particles with a self-consistent electric field.
The plasma is assumed to be quasi-neutral and composed of electrons and immobile background ions. 
The normalized Vlasov--Poisson equations are
\begin{align}
    \frac{\mathrm{d}}{\mathrm{d} t} f(x, v, t) \coloneqq \left[\partial_{t} + v \partial_{x}+ \partial_{x} \phi(x, t)\partial_{v}\right] f(x, v, t)  &= 0, \label{vlasov-continuum}\\
    -\partial_{x}^{2} \phi(x, t) &= 1 - \int_{-\infty}^{\infty} f(x,v, t) \mathrm{d}v, \label{poisson-continuum}
\end{align}
where $f(x, v, t)$ is the electron distribution function and $\phi(x, t)$ is the electrostatic potential. We consider an unbounded velocity coordinate $v \in \mathbb{R}$, a periodic spatial coordinate $x \in [0, \ell]$, where $\ell$ is the length of the spatial domain, and time $t \geq 0$. 

\paragraph{Normalization}{
The quantities above are normalized as follows:
\begin{equation*}
   t \coloneqq t_{d} \omega_{pe}, \qquad x \coloneqq \frac{x_{d}}{\lambda_{D}}, \qquad v \coloneqq \frac{v_{d}}{v_{te}}, \qquad f \coloneqq f_{d}\frac{v_{te}}{n_{e}},  \qquad \phi \coloneqq \phi_{d} \frac{e \lambda_{D}}{T_{e}}, 
\end{equation*}
where the subscript `$d$' indicates the dimensional quantities, $e$ is the positive elementary charge, $\omega_{pe} \coloneqq \sqrt{4 \pi e^2 n_{e}/m_{e}}$ is the electron plasma frequency, $m_{e}$ is the electron mass, $n_{e}$ is the reference electron density, $v_{te} \coloneqq \sqrt{T_{e}/m_{e}}$ is the electron thermal velocity, $T_{e}$ is a reference electron temperature, and $\lambda_{D} \coloneqq v_{te}/\omega_{pe}$ is the electron Debye length.}

\subsection{Asymmetrically weighted Hermite basis functions}\label{sec:hermite}
The AW Hermite basis functions are defined as
\begin{align}
    \psi_{n}(v;u, \alpha) &\coloneqq (\pi 2^n n!)^{-\frac{1}{2}} H_{n}\left(\frac{v-u}{\alpha}\right) \exp{\left(-\left[\frac{v-u}{\alpha}\right]^2\right)}, \label{Hermite-basis-function}\\
    \psi^{n}(v; u, \alpha) &\coloneqq (2^n n!)^{-\frac{1}{2}} H_{n}\left(\frac{v -u}{\alpha}\right), \label{Hermite-basis-function-dual}
\end{align}
where $H_{n}$ is the $n$th \textit{physicist's} Hermite polynomial~\cite[\S 22.3.10]{abramowitz_1964_math}.
The two Hermite parameters $u \in \mathbb{R}$ and $\alpha \in \mathbb{R}_{>0}$ are the velocity shifting and scaling parameters, respectively. 
Adjusting these two parameters can significantly improve the Hermite spectral convergence~\cite{schumer_1998_jcp, pagliantini_2023_jcp, shao_2025_adaptive}.
The orthogonality and recursive properties of the Hermite basis~\cite[\S 22.2--22.8]{abramowitz_1964_math} are 
\begin{align}
    \int_{-\infty}^{\infty} \psi_{n}\psi^{m} \mathrm{d} v &= \alpha \delta_{n, m}, \label{orthogonality-AW}\\
    \frac{\mathrm{d}\psi_{n}}{\mathrm{d}v}  &= - \frac{\sqrt{2(n+1)}}{\alpha}\psi_{n+1}, \label{recursive-AW-1}\\
    v\psi_{n} &= \alpha \sqrt{\frac{n+1}{2}} \psi_{n+1} + \alpha \sqrt{\frac{n}{2}} \psi_{n-1} + u \psi_{n}, \label{recursive-AW-2}
\end{align}
where $\delta_{m, n}$ denotes the Kronecker delta function.
%

\subsection{Legendre basis functions}\label{sec:legendre}
The Legendre basis functions are defined as 
\begin{equation}\label{legendre_basis}
    \xi_{n}(v; v_{a}, v_{b}) \coloneqq \sqrt{2n + 1} L_{n}\left(\frac{2v - [v_{a} + v_{b}]}{v_{b} - v_{a}}\right),
\end{equation}
where $L_{n}(z)$ is the $n$th Legendre polynomial~\cite[\S 22.3.8]{abramowitz_1964_math} on the interval $z \in [-1, 1]$ such that we remap the Legendre polynomials on the range $v \in [v_{a}, v_{b}]$ via a linear transformation introduced in~\citet{manzini_2016_jcp}.
Some useful orthogonality and recursive properties~\cite[\S 22.2-22.8]{abramowitz_1964_math} are as follows:  
\begin{align}
    \int_{v_{a}}^{v_{b}} \xi_{n} \xi_{m} \mathrm{d} v &= (v_{b}-v_{a}) \delta_{n, m},\label{orthogonality-legendre}\\
    \frac{\mathrm{d} \xi_{n}}{\mathrm{d} v} &= \sum_{i=0}^{n-1} \sigma_{n, i} \xi_{i}, \label{recursive-legendre-1} \\
    \sigma_{n, i} &\coloneqq \begin{cases}
        0 & n-i \text{ is even,}\\
        \frac{2\sqrt{(2n + 1)(2i + 1)}}{v_{b} - v_{a}} & n-i \text{ is odd,}
    \end{cases}\nonumber \\
    v \xi_{n} &= \sigma_{n+1} \xi_{n+1} + \sigma_{n} \xi_{n-1} + \bar{\sigma} \xi_{n}.\label{recursive-legendre-2}
\end{align}

\subsection{Mixed Hermite--Legendre method}\label{sec:mixed_method_derived}
Consider the following decomposition of the electron distribution function
\begin{equation}\label{decomposition_of_f}
    f(x, v, t)= f_{0}(x, v, t)+ \delta f(x, v, t),
\end{equation}
where $f_{0}$ and $\delta f$ are the parts of the distribution function near and far from Maxwellian (or local thermal equilibrium), respectively. 
The decomposition in Eq.~\eqref{decomposition_of_f} is not \underline{unique}. 
Inserting the decomposition in Eq.~\eqref{decomposition_of_f} in the Vlasov equation~\eqref{vlasov-continuum} results in
\begin{equation}\label{f0-df-evolution}
    \frac{\mathrm{d} f_{0}}{\mathrm{d} t} = - \frac{\mathrm{d} \delta f}{\mathrm{d} t}.
\end{equation}
The static method in~\citet{chapurin_2024_pop} imposes $\frac{\mathrm{d} f_{0}}{\mathrm{d} t} = \frac{\mathrm{d} \delta f}{\mathrm{d} t} = 0$, ensuring that both $f_{0}$ and $\delta f$ individually satisfy the Vlasov equation~\eqref{vlasov-continuum}. In contrast, the proposed mixed method dynamically redistributes information between $f_{0}$ and $\delta f$, as we demonstrate next.

We begin by expanding $f_{0}$ in velocity via the AW Hermite basis:
\begin{equation}\label{f0-expansion}
    f_{0}(x, v, t) = \sum_{n=0}^{N_{H}-1} C_{n}(x, t) \psi_{n}(v; \alpha, u),
\end{equation}
where $\psi_{n}$ is the $n$th AW Hermite basis function defined in Eq.~\eqref{Hermite-basis-function}.
After inserting the $f_{0}$ expansion~\eqref{f0-expansion} in the Vlasov equation~\eqref{vlasov-continuum} and employing the AW Hermite basis recursive identities in Eqns.~\eqref{recursive-AW-1}--\eqref{recursive-AW-2}, we get
\begin{align}\label{inserting_deocomposition}
    \frac{\mathrm{d} f_{0}}{\mathrm{d} t} &= \sum_{n=0}^{N_{H}-1} \left[\psi_{n} \partial_{t} C_{n}  + v \psi_{n}\partial_{x}C_{n} +   C_{n} \frac{\mathrm{d}\psi_{n}}{\mathrm{d}v} \partial_{x} \phi\right] \nonumber\\
    &=\sum_{n=0}^{N_{H}-1} \left[\psi_{n}\partial_{t} C_{n}  +  \left[\alpha \sqrt{\frac{n+1}{2}} \psi_{n+1} +  \alpha\sqrt{\frac{n}{2}} \psi_{n-1} + u \psi_{n}\right] \partial_{x}C_{n}  - \frac{\sqrt{2(n+1)}}{\alpha} C_{n} \psi_{n+1} \partial_{x} \phi \right] \\
    &= - \frac{\mathrm{d} \delta f}{\mathrm{d} t}. \nonumber
\end{align}
Multiplying Eq.~\eqref{inserting_deocomposition} by the AW Hermite dual basis function $\psi^{n}(v; u, \alpha)$ defined in Eq.~\eqref{Hermite-basis-function-dual}, employing the orthogonality relation of the AW Hermite basis in Eq.~\eqref{orthogonality-AW}, and integrating with respect to velocity, gives rise to
\begin{align}\label{f0-evolution-equation}
 \partial_{t} C_{n} + \alpha \sqrt{\frac{n+1}{2}} \partial_{x} C_{n+1} + \alpha \sqrt{\frac{n}{2}} \partial_{x} C_{n-1} + u \partial_{x} C_{n} - \frac{\sqrt{2n}}{\alpha} C_{n-1} \partial_{x} \phi = - \frac{1}{\alpha} \int \frac{\mathrm{d} \delta f}{\mathrm{d} t} \psi^{n} \mathrm{d} v,
\end{align}
for $n = 0, 1, \ldots, N_{H}-1$. 
Next, we insert Eq.~\eqref{f0-evolution-equation} in Eq.~\eqref{inserting_deocomposition}. It follows that 
\begin{equation}\label{df-evolution-equation}
 \frac{\mathrm{d} \delta f}{\mathrm{d} t}  = \frac{1}{\alpha}  \left[\sum_{n=0}^{N_{H} -1}\int \frac{\mathrm{d} \delta f}{\mathrm{d} t} \psi^{n} \mathrm{d} v \psi_{n} \right]- \alpha \sqrt{\frac{N_{H}}{2}} \psi_{N_{H}} \partial_{x} C_{N_{H}-1}+ \alpha \sqrt{\frac{N_{H}}{2}} \psi_{N_{H}-1} \partial_{x} \cancelto{0}{C_{N_{H}}}+\frac{\sqrt{2N_{H}}}{\alpha} \psi_{N_{H}} C_{N_{H} -1} \partial_{x} \phi.
\end{equation}
In the above, we assume $C_{N_{H}} = 0$, which is also known as \textit{closure by truncation}. 
We highlight that the continuous Vlasov equation~\eqref{vlasov-continuum} and Eqns.~\eqref{f0-evolution-equation}--\eqref{df-evolution-equation} are equivalent, since so far we have only enforced the description of $f_{0}$.

The mixed method is obtained by imposing 
\begin{equation}
     \int_{-\infty}^{\infty} \frac{\mathrm{d} \delta f}{\mathrm{d} t} \psi^{n} \mathrm{d} v = 0, \qquad \text{for} \Hquad  n = 0,1, \ldots, N_{H}-1.
     \label{constraint}
\end{equation}
Other constraints are of course possible (and will be explored in future work) and this one was chosen for the simplicity of the resulting coupled equations, where, as can be seen below, only the equation for $\delta f$ includes an explicit coupling with the equation for $f_0$.
From the Poisson equation~\eqref{poisson-continuum} and Eqns.~\eqref{f0-evolution-equation}--\eqref{df-evolution-equation}, the constraint above leads to the following evolution equations
\begin{align}
    \partial_{t} C_{n} &+ \alpha \sqrt{\frac{n+1}{2}} \partial_{x} C_{n+1} + \alpha \sqrt{\frac{n}{2}} \partial_{x} C_{n-1} + u \partial_{x} C_{n} - \frac{\sqrt{2n}}{\alpha} C_{n-1}  \partial_{x} \phi=0, \label{mixed_method_1_a}\\
    \frac{\mathrm{d} \delta f}{\mathrm{d} t}  &= - \alpha \sqrt{\frac{N_{H}}{2}} \left[\partial_{x}  - \frac{2}{\alpha^2}\partial_{x} \phi  \right] \psi_{N_{H}} C_{N_{H} -1}, \label{mixed_method_1_b}\\
    -\partial_{x}^{2} \phi &= 1 - \alpha C_{0} - \int_{-\infty}^{\infty} \delta f \mathrm{d}v. \label{mixed_method_1_c}
\end{align}
We use the Legendre basis to discretize $\delta f$ as follows
\begin{align}\label{df-expansion}
    \delta f(x, v, t) &= \sum_{n=0}^{N_{L}-1} B_{n}(x, t) \xi_{n}(v; v_{a}, v_{b}),
\end{align}
where $\xi_{n}$ is the $n$th Legendre basis function defined in Eq.~\eqref{legendre_basis}. 
Inserting the Legendre spectral expansion in Eq.~\eqref{df-expansion} in Eq.~\eqref{mixed_method_1_b}, multiplying by $\xi_{m}(v; v_{a}, v_{b})$, employing the Legendre recursive and orthogonality identities in Eqns.~\eqref{orthogonality-legendre}--\eqref{recursive-legendre-2}, and integrating with respect to velocity, results in
\begin{align*}
    \partial_{t} B_{m} + \sigma_{m} \partial_{x} B_{m-1}+ \sigma_{m+1}\partial_{x} B_{m+1}  &+ \bar{\sigma}\partial_{x} B_{m} - \partial_{x} \phi \left[ \sum_{i=0}^{m-1} \sigma_{m, i} B_{i}  - \gamma_{m} \delta_{v}[\delta f \xi_{m}]_{v_{a}}^{v_{b}} \right] \\
    &= -\frac{\alpha}{v_{b}-v_{a}} \mathcal{J}_{N_{H}, m}\sqrt{\frac{N_{H}}{2}} \left[\partial_{x}  - \frac{2}{\alpha^2} \partial_{x} \phi  \right] C_{N_{H} -1} ,
\end{align*}
for $m=0, 1, \ldots N_{L}-1$, where 
\begin{align}
    \mathcal{J}_{n, m} &\coloneqq \int_{v_{a}}^{v_{b}} \psi_{n}\xi_{m} \mathrm{d} v, \nonumber\\
    \delta_{v}[\delta f \xi_{m}]_{v_{a}}^{v_{b}} &\coloneqq \frac{\delta f(x, v_{b}, t) \xi_{m}(v_{b}) - \delta f(x, v_{a}, t) \xi_{m}(v_{a})}{v_{b} - v_{a}}, \nonumber \\
    \gamma_{m} &\coloneqq \begin{cases}
        \gamma, \Hquad m\geq 3, \\
        0, \Hquad m < 3.
    \end{cases}\nonumber\\
    \bar{\sigma} &\coloneqq \frac{v_{a} + v_{b}}{2},\nonumber\\
    \sigma_{n} &\coloneqq \begin{cases}
        0 & n=0,\\
        \frac{v_{b}-v_{a}}{2} \frac{n}{\sqrt{(2n+1)(2n-1)}} & n \geq 1.
    \end{cases}\nonumber
\end{align}
In the above, we follow the work by~\citet{manzini_2016_jcp}, which introduces the boundary conditions $\delta f(x, v_{a}, t) = \delta f(x, v_{b}, t) = 0$ in weak form via the penalty coefficient $\gamma_{m}$. 
Similar to~\citet{manzini_2016_jcp}, we apply the penalty term to all Legendre modes except the first three equations to improve the method's conservation properties. 
To close the set of equations, we set $B_{N_{L}} = 0$, i.e., \textit{closure by truncation}.

\begin{figure}
    \centering
    \includegraphics[width=0.7\linewidth]{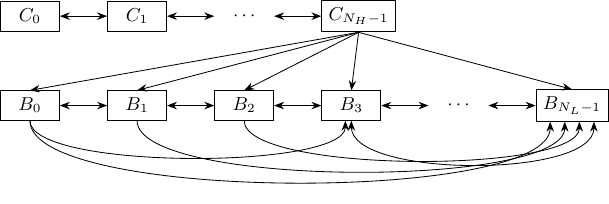}
    \caption{Mixed method equation coupling represented as a graph network, with Legendre coefficients $B_m$ for $\delta f$ and Hermite coefficients $C_n$ for $f_0$. Removing the coupling between the final Hermite coefficient and the first three Legendre coefficients ensures conservation of mass, momentum, and energy independent of the spectral parameters; we discuss this further in section~\ref{sec:conservation_laws}. }
    \label{fig:mixed_method_illustration}
\end{figure}

In summary, the mixed method system of equations is given by 
\begin{align}
    \partial_{t} C_{n} &+ \alpha \sqrt{\frac{n+1}{2}} \partial_{x} C_{n+1} + \alpha \sqrt{\frac{n}{2}} \partial_{x} C_{n-1} + u \partial_{x} C_{n} - \frac{\sqrt{2n}}{\alpha} C_{n-1}  \partial_{x} \phi=0, \label{mixed_method_1_a_discrete}\\
    \partial_{t} B_{m} &+ \sigma_{m} \partial_{x} B_{m-1}+ \sigma_{m+1}\partial_{x} B_{m+1}  + \bar{\sigma}\partial_{x} B_{m} - \partial_{x} \phi \left[ \sum_{i=0}^{m-1} \sigma_{m, i} B_{i}  - \gamma_{m} \delta_{v}[\delta f \xi_{m}]_{v_{a}}^{v_{b}} \right] \label{mixed_method_1_b_discrete}\\
    &= -\frac{\alpha}{v_{b}-v_{a}} \mathcal{J}_{N_{H}, m} \sqrt{\frac{N_{H}}{2}}  \left[\partial_{x}  - \frac{2}{\alpha^2} \partial_{x} \phi  \right] C_{N_{H} -1} , \nonumber\\
    -\partial_{x}^{2} \phi &= 1 - \alpha C_{0} - (v_{b} - v_{a}) B_{0}. \label{mixed_method_1_c_discrete}
\end{align}
Analogous to a chain of spring-mass-dampers, each coefficient of order $n$ interacts with its nearest neighbors ($n\pm1$), except that the highest Hermite coefficient couples explicitly to all Legendre coefficients, providing a direct one-way coupling of the two representations, see Figure~\ref{fig:mixed_method_illustration} for a graph illustration of the spectral coefficient coupling.
Moreover, $\delta f$ and $f_{0}$ are also indirectly coupled through the Poisson equation~\eqref{mixed_method_1_c_discrete}.
The mixed method is most advantageous when non-Maxwellian features are confined to a relatively small region in velocity space. In such cases, the Legendre representation is defined over a smaller domain and can achieve higher accuracy. This follows from the asymptotic relation of the Legendre basis function~\cite[\S 22.15]{abramowitz_1964_math}:
\begin{equation*}
    \lim_{n \to \infty} \xi_{n}(v; v_{a}, v_{b}) \propto J_{0}\left(n y\right) \sim y^{-\frac{1}{2}}\cos\left(n y - \frac{\pi}{4}\right), 
    \qquad \text{with} \quad 
    y \coloneqq \arccos\left(\frac{2v - (v_{a} + v_{b})}{v_{b} - v_{a}}\right),
\end{equation*}
where $J_{0}$ is the zeroth-order Bessel function of the first kind~\cite[\S 9.2.1]{abramowitz_1964_math}. 
Therefore, the shortest wavelength $\lambda_{L}$ in velocity space associated with the Legendre spectral expansion of order $N_{L}$ is 
\begin{equation*}
    \lambda_{L} \approx \frac{\pi (v_{b} - v_{a})}{N_{L}},
\end{equation*}
such that reducing the domain size $v_{b} - v_{a}$ decreases the minimum resolved velocity wavelength, improving the accuracy of the Legendre representation of $\delta f$. Resolving local features in velocity space is in general harder for Hermite functions, since these are global functions defined over the whole velocity space.

\subsection{Artificial collisional operator}\label{sec:collisions}
Collisionless plasmas can develop increasingly fine velocity-space structures, leading to phase-space filamentation~\cite{grant_1967_pof, schumer_1998_jcp, issan_2024_pop}.
Filamentation can cause spectral methods to develop numerical instabilities and recurrence, where the solution is artificially periodic in time. 
To control filamentation effects, we modify Eqns.~\eqref{mixed_method_1_a_discrete}--\eqref{mixed_method_1_b_discrete} by introducing an artificial collisional operator based on the Lenard-Bernstein collisional operator~\cite{lenard_1958_pr} in the continuum to the right-hand side:
\begin{align*}
    \mathcal{C}(B_{m}) &= - \nu_{L} \frac{m(m-1)(m-2)}{(N_{L} -1)(N_{L}-2)(N_{L}-3)} B_{m},\\
    \mathcal{C}(C_{n}) &= - \nu_{H} \frac{n(n-1)(n-2)}{(N_{H} -1)(N_{H}-2)(N_{H}-3)} C_{n},
\end{align*}
where $\nu_{L}, \nu_{H} \in \mathbb{R}_{\geq 0}$ are the artificial collisional rates. 
The collisional operator purposefully only acts on higher-order coefficients to preserve the method's conservation properties, since mass, momentum, and energy are described by the first three Hermite and Legendre coefficients~\cite{camporeale_2016_cpc, manzini_2016_jcp}. 
Note that in order to fully take advantage of the mixed method automated adaptivity of $\delta f$ from $f_{0}$, it is important to keep $\nu_{H} \to 0$, since the mixed method relies on the last Hermite moment to transfer information to $\delta f$.

\section{Conservation laws}\label{sec:conservation_laws}
In this section, we derive the conservation properties of the mixed method, including total mass in section~\ref{sec:mass_conservation}, total momentum in section~\ref{sec:momentum_conservation}, and total energy in section~\ref{sec:energy_conservation}. 
Lastly, we summarize the conservation properties and discuss practical considerations in section~\ref{sec:conservation_summary}.

\subsection{Conservation of mass}\label{sec:mass_conservation}
The total electron mass is defined as 
\begin{equation}\label{mass-definition}
    \mathcal{M}(t) \coloneqq  \int_{0}^{\ell} \int_{-\infty}^{\infty} f \mathrm{d} v \mathrm{d} x =  \alpha \int_{0}^{\ell} C_{0} \mathrm{d} x + (v_{b} - v_{a}) \int_{0}^{\ell} B_{0} \mathrm{d} x.
\end{equation}
By taking the time derivative of Eq.~\eqref{mass-definition} and inserting Eqns.~\eqref{mixed_method_1_a_discrete}--\eqref{mixed_method_1_b_discrete}, it follows that 
\begin{align*}
    \frac{\mathrm{d}\mathcal{M}}{\mathrm{d} t} &= \alpha \int_{0}^{\ell} \partial_{t} C_{0} \mathrm{d} x + (v_{b} - v_{a}) \int_{0}^{\ell} \partial_{t} B_{0} \mathrm{d} x \\
    &= \alpha \left[-\frac{\alpha}{\sqrt{2}} \cancelto{0}{\int_{0}^{\ell} \partial_{x} C_{1} \mathrm{d} x}  - u \cancelto{0}{\int_{0}^{\ell} \partial_{x} C_{0} \mathrm{d} x}\right] + (v_{b} - v_{a}) \left[ - \sigma_{1} \cancelto{0}{\int_{0}^{\ell} \partial_{x} B_{1} \mathrm{d} x} - \bar{\sigma}  \cancelto{0}{ \int_{0}^{\ell} \partial_{x} B_{0} \mathrm{d} x}\right] \\
    &-\alpha \sqrt{\frac{N_{H}}{2}} \mathcal{J}_{N_{H}, 0} \cancelto{0}{\int_{0}^{\ell} \partial_{x}  C_{N_{H}-1} \mathrm{d} x} + \frac{\sqrt{2N_{H}}}{\alpha} \mathcal{J}_{N_{H}, 0} \int_{0}^{\ell}  C_{N_{H}-1} \partial_{x} \phi\mathrm{d} x\\
    &= \frac{\sqrt{2N_{H}}}{\alpha} \mathcal{J}_{N_{H}, 0} \int_{0}^{\ell}  C_{N_{H}-1} \partial_{x} \phi\mathrm{d} x.
\end{align*}
The crossed-out terms above cancel out due to periodic spatial boundary conditions. 
The integral $\mathcal{J}_{N_{H}, 0}$ can be simplified via even/odd symmetric properties of Hermite polynomials, such that 
\begin{equation}\label{J0}
    \mathcal{J}_{N_{H}, 0} \coloneqq \int_{v_{a}}^{v_{b}} \psi_{N_{H}} \xi_{0} \mathrm{d} v \propto \int_{v_{a}}^{v_{b}} \mathcal{H}_{N_{H}}\left(\frac{v-u}{\alpha}\right)\exp{\left(-\left(\frac{v-u}{\alpha}\right)^2\right)} \mathrm{d} v = \begin{cases}
    0, \Hquad u= \bar{\sigma} \text{ and } N_{H} \text{ is odd},\\
    0, \Hquad -v_{a} = v_{b} \to \infty  \text{ and } N_{H} \geq 1, \\
    \text{non-zero}, \Hquad \text{else}.
    \end{cases} 
\end{equation}
The second case for which $-v_{a} = v_{b} \to \infty$ holds since $\xi_{0} \propto \psi^{0}$. 
Therefore, the mixed method total mass is conserved if $N_{H}$ is odd and $u = \bar{\sigma} \coloneqq \frac{v_{a} + v_{b}}{2}$ or if $-v_{a} = v_{b} \to \infty$ and $N_{H} \geq 1$.


\subsection{Conservation of momentum}\label{sec:momentum_conservation}
The total electron momentum is defined as 
\begin{align}\label{momentum-definiton-mm1}
    \mathcal{P}(t) &\coloneqq \int_{0}^{\ell} \int_{-\infty}^{\infty} v f\mathrm{d} v \mathrm{d} x = \int_{0}^{\ell} \left[ \frac{\alpha^2}{\sqrt{2}}  C_{1} + u \alpha C_{0} + (v_{b} - v_{a}) \sigma_{1} B_{1} + \bar{\sigma} (v_{b}-v_{a}) B_{0} \right] \mathrm{d} x. 
\end{align}
By taking the time derivative of Eq.~\eqref{momentum-definiton-mm1} and inserting Eqns.~\eqref{mixed_method_1_a_discrete}--\eqref{mixed_method_1_c_discrete}, it follows that 
\begin{align*}
    \frac{\mathrm{d}\mathcal{P}}{\mathrm{d} t} &=  \underbrace{\frac{\alpha^2}{\sqrt{2}}   \int_{0}^{\ell}\partial_{t} C_{1} \mathrm{d} x}_{(\mathrm{I})} + u \alpha \cancelto{0}{ \int_{0}^{\ell} \partial_{t}  C_{0} \mathrm{d} x} + \underbrace{(v_{b} - v_{a}) \sigma_{1} \int_{0}^{\ell} \partial_{t} B_{1} \mathrm{d} x}_{(\mathrm{II})} + \underbrace{\bar{\sigma} (v_{b}-v_{a}) \int_{0}^{\ell} \partial_{t} B_{0} \mathrm{d} x}_{(\mathrm{III})}
\end{align*}
We proceed by simplifying each term 
\begin{alignat*}{2}
    (\mathrm{I}) &\hspace{50pt}\frac{\alpha^2}{\sqrt{2}}  \int_{0}^{\ell}\partial_{t} C_{1} \mathrm{d} x &&= -(v_{b} - v_{a}) \int_{0}^{\ell} B_{0} \partial_{x} \phi  \mathrm{d} x\\
     (\mathrm{II}) &\qquad (v_{b} - v_{a}) \sigma_{1} \int_{0}^{\ell} \partial_{t} B_{1} \mathrm{d} x &&= (v_{b} - v_{a}) \int_{0}^{\ell} B_{0} \partial_{x} \phi \mathrm{d} x + \sigma_{1} \frac{\sqrt{2N_{H}} }{\alpha} \mathcal{J}_{H_{H}, 1}  \int_{0}^{\ell}C_{N_{H} -1} \partial_{x} \phi \mathrm{d} x\\
     (\mathrm{III}) &\qquad  \bar{\sigma} (v_{b}-v_{a}) \int_{0}^{\ell} \partial_{t} B_{0} \mathrm{d} x &&= \bar{\sigma}  \frac{\sqrt{2N_{H}}}{\alpha} \mathcal{J}_{N_{H}, 0} \int_{0}^{\ell} C_{N_{H}-1} \partial_{x} \phi \mathrm{d} x
\end{alignat*}
Then, we get 
\begin{equation*}
    \frac{\mathrm{d}\mathcal{P}}{\mathrm{d} t} = \frac{\sqrt{2N_{H}}}{\alpha} \left[\bar{\sigma} \mathcal{J}_{N_{H}, 0} + \sigma_{1}  \mathcal{J}_{N_{H}, 1}\right]\int_{0}^{\ell} C_{N_{H}-1}\partial_{x} \phi \mathrm{d} x.
\end{equation*}
The integral $\mathcal{J}_{N_{H}, 1}$ can be simplified via even/odd properties of Hermite polynomials, such that 
\begin{align}\label{J1}
    \mathcal{J}_{N_{H}, 1} &\coloneqq \int_{v_{a}}^{v_{b}} \psi_{N_{H}} \xi_{1} \mathrm{d} v \propto \int_{v_{a}}^{v_{b}} \mathcal{H}_{N_{H}}\left(\frac{v-u}{\alpha}\right) \exp{\left(-\left(\frac{v-u}{\alpha}\right)^2\right)}  L_{1}\left(\frac{2v - [v_{a} + v_{b}]}{v_{b} - v_{a}}\right) \mathrm{d} v \\
    &= \begin{cases}
    0, \Hquad  u = \bar{\sigma} \text{ and } N_{H} \text{ is even},\\
    0, \Hquad -v_{a} = v_{b} \to \infty \text{ and } N_{H} \geq 2,\\
    \text{non-zero}, \Hquad \text{else}.
    \end{cases}\nonumber
\end{align}
The second case is because $\xi_{1} \propto c_{1} \psi^{0} + c_{2} \psi^{1}$, where $c_{1}, c_{2} \in \mathbb{R}$ are constants. Thus, total momentum is conserved if $-v_{a} = v_{b}$ (since then $\bar{\sigma} = 0$), $u=0$, and $N_{H}$ is even or if $-v_{a} = v_{b} \to \infty$ and $N_{H} \geq 2$.

\subsection{Conservation of energy}\label{sec:energy_conservation}
The total energy is defined as 
\begin{align}
    \mathcal{E}(t) &\coloneqq \mathcal{E}^{e}_{\mathrm{kin}}(t) + \mathcal{E}^{i}_{\mathrm{kin}} + \mathcal{E}_{\mathrm{pot}}(t), \nonumber\\
    \mathcal{E}_{\mathrm{kin}}^{e}(t) &\coloneqq \frac{1}{2} \int_{0}^{\ell} \int_{-\infty}^{\infty} v^2 f \mathrm{d} v \mathrm{d} x =\frac{\alpha}{2} \int_{0}^{\ell} \left[\frac{\alpha^2}{\sqrt{2}} C_{2} + \sqrt{2} u \alpha C_{1} + \left[\frac{\alpha^2}{2} + u^2\right] C_{0} \right]\mathrm{d} x \label{kinetic-energy-definition-mm1}\\
    &+ \frac{[v_{b}-v_{a}]}{2} \int_{0}^{\ell}\left[\sigma_{2} \sigma_{1} B_{2} + 2\sigma_{1} \bar{\sigma} B_{1} + \left[\sigma_{1}^2 + \sigma_{0}^2 + \bar{\sigma}^2\right] B_{0} \right] \mathrm{d} x, \nonumber\\
    \mathcal{E}_{\mathrm{pot}}(t) &\coloneqq \frac{1}{2} \int_{0}^{\ell} E^2 \mathrm{d} x, \label{potential-energy-definition-mm1}
\end{align}
where $\mathcal{E}^{e}_{\mathrm{kin}}(t)$ is the electron kinetic energy, $\mathcal{E}^{i}_{\mathrm{kin}}$ is the ion kinetic energy, and $\mathcal{E}_{\mathrm{pot}}(t)$ is the potential energy. 
By taking the time derivative of Eq.~\eqref{kinetic-energy-definition-mm1} and inserting Eqns.~\eqref{mixed_method_1_a_discrete}--\eqref{mixed_method_1_b_discrete}, we get
\begin{align}
    \frac{\mathrm{d}\mathcal{E}_{\mathrm{kin}}^{e}}{\mathrm{d}t} &=\underbrace{\frac{\alpha^3}{2\sqrt{2}}  \int_{0}^{\ell}\partial_{t} C_{2}\mathrm{d} x}_{(\mathrm{I})} + \underbrace{\frac{u \alpha^2}{\sqrt{2}} \int_{0}^{\ell} \partial_{t} C_{1}\mathrm{d} x}_{(\mathrm{II})} + \frac{\alpha}{2}\left[\frac{\alpha^2}{2} + u^2\right] \cancelto{0}{ \int_{0}^{\ell} \partial_{t} C_{0} \mathrm{d} x} \nonumber\\
     &+ \underbrace{\frac{[v_{b}-v_{a}]\sigma_{2} \sigma_{1} }{2} \int_{0}^{\ell} \partial_{t} B_{2}\mathrm{d} x}_{(\mathrm{III})} + \underbrace{ [v_{b} - v_{a}] \sigma_{1} \bar{\sigma} \int_{0}^{\ell} \partial_{t} B_{1}\mathrm{d} x}_{(\mathrm{IV})} + \underbrace{\frac{[v_{b}-v_{a}]\left[\sigma_{1}^2 + \sigma_{0}^2 + \bar{\sigma}^2\right]}{2} \int_{0}^{\ell} \partial_{t} B_{0}\mathrm{d} x}_{(\mathrm{V})} \label{kin-energy-variation}
\end{align}
We proceed by simplifying each term as follows
\begin{alignat*}{2}
    (\mathrm{I}) &\hspace{100pt}\frac{\alpha^3}{2\sqrt{2}} \int_{0}^{\ell} \partial_{t} C_{2} \mathrm{d} x &&= \frac{\alpha^2}{\sqrt{2}} \int_{0}^{\ell}  C_{1} \partial_{x} \phi \mathrm{d} x\\
    (\mathrm{II}) &\hspace{100pt}\frac{u\alpha^2}{\sqrt{2}} \int_{0}^{\ell} \partial_{t} C_{1} \mathrm{d} x &&= u \alpha \int_{0}^{\ell}C_{0} \partial_{x} \phi \mathrm{d} x\\
    (\mathrm{III}) &\hspace{60pt} \frac{[v_{b}-v_{a}]\sigma_{2} \sigma_{1}}{2} \int_{0}^{\ell} \partial_{t} B_{2} \mathrm{d} x &&= \sigma_{1}[v_{b} - v_{a}] \int_{0}^{\ell} B_{1} \partial_{x} \phi\mathrm{d} x + \frac{\sqrt{N_{H}} \sigma_{1} \sigma_{2}}{\sqrt{2} \alpha} \mathcal{J}_{N_{H}, 2} \int_{0}^{\ell} C_{N_{H} - 1} \partial_{x} \phi\mathrm{d} x\\
    (\mathrm{IV}) &\hspace{50pt}\qquad [v_{b} - v_{a}] \sigma_{1} \bar{\sigma} \int_{0}^{\ell} \partial_{t} B_{1} \mathrm{d} x &&= \bar{\sigma}[v_{b} - v_{a}] \int_{0}^{\ell}  B_{0} \partial_{x} \phi\mathrm{d} x + \frac{\sqrt{2N_{H}} \sigma_{1} \bar{\sigma}}{\alpha} \mathcal{J}_{N_{H}, 1} \int_{0}^{\ell} C_{N_{H} - 1} \partial_{x} \phi \mathrm{d} x\\
    (\mathrm{V}) &\qquad \frac{[v_{b}-v_{a}][\sigma_{1}^2 + \sigma_{0}^2 + \bar{\sigma}^2]}{2}  \int_{0}^{\ell} \partial_{t} B_{0} \mathrm{d} x &&= \frac{\sqrt{N_{H}}[\sigma_{1}^2 + \sigma_{0}^2 + \bar{\sigma}^2]}{\sqrt{2}\alpha}  \mathcal{J}_{N_{H}, 0} \int_{0}^{\ell}  C_{N_{H}-1} \partial_{x} \phi\mathrm{d} x 
\end{alignat*}
Similarly, the time derivative of the potential energy is 
\begin{align}
    \frac{\mathrm{d}\mathcal{E}_{\mathrm{pot}}}{\mathrm{d} t} &= -\int_{0}^{\ell} [\sigma_{1} [v_{b} - v_{a}] B_{1} + \bar{\sigma} [v_{b} - v_{a}]  B_{0}+ \frac{\alpha^2}{\sqrt{2}}  C_{1} + u\alpha C_{0}] \partial_{x} \phi \mathrm{d} x \label{potential-energy-variation}
\end{align}
Therefore, by adding Eq.~\eqref{potential-energy-variation} and Eq.~\eqref{kin-energy-variation}, we get 
\begin{align*}
   \frac{\mathrm{d}\mathcal{E}}{\mathrm{d} t} = \frac{\mathrm{d}\mathcal{E}_{\mathrm{kin}}^{e}}{\mathrm{d}t} + \frac{\mathrm{d}\mathcal{E}_{\mathrm{pot}}}{\mathrm{d} t} =  \frac{1}{\alpha}\sqrt{\frac{N_{H}}{2}} \left[\sigma_{1}\sigma_{2} \mathcal{J}_{N_{H}, 2}+ 2 \sigma_{1} \bar{\sigma} \mathcal{J}_{N_{H}, 1} + [\sigma_{1}^2 + \sigma_{0}^2 + \bar{\sigma}^2] \mathcal{J}_{N_{H}, 0} \right]\int_{0}^{\ell} C_{N_{H}-1}\partial_{x} \phi\mathrm{d} x.
\end{align*}
The integral $\mathcal{J}_{N_{H}, 2}$ can be simplified via even/odd symmetric properties of Hermite polynomials, such that 
\begin{align}\label{J2}
    \mathcal{J}_{N_{H}, 2} &\coloneqq \int_{v_{a}}^{v_{b}} \psi_{N_{H}}\xi_{2} \mathrm{d} v \propto \int_{v_{a}}^{v_{b}} \mathcal{H}_{N_{H}}\left(\frac{v-u}{\alpha}\right) \exp{\left(-\left(\frac{v-u}{\alpha}\right)^2\right)} L_{2}\left(\frac{2v - [v_{a} + v_{b}]}{v_{b} - v_{a}}\right)\mathrm{d} v\\ \nonumber
    &= \begin{cases}
    0, \Hquad u = \bar{\sigma} \text{ and } N_{H} \text{ is odd}\\
    0, \Hquad -v_{a} = v_{b} \to \infty \text{ and } N_{H} \geq 3\\
    \text{non-zero}, \Hquad \text{else}
    \end{cases} 
\end{align}
The second case holds because $\xi_{2} \propto c_{1} \psi^{0} + c_{2} \psi^{1} + c_{3} \psi^{2}$, where $c_{1}, c_{2}, c_{3} \in \mathbb{R}$ are constants. 
Thus, energy is conserved if $-v_{a} = v_{b}$ (since then $\bar{\sigma} = 0$), $u= 0$, and $N_{H}$ is odd (since then $\mathcal{J}_{N_{H}, 0}=\mathcal{J}_{N_{H}, 2} = 0$) or if $-v_{a} = v_{b} \to \infty \text{ and } N_{H} \geq 3$.


\subsection{Conservation properties summary}\label{sec:conservation_summary}
\begin{figure}
    \centering
    \begin{subfigure}[b]{0.51\textwidth}
        \centering 
        \caption{$|\mathcal{J}_{N_{H}, 0}|$}
        \includegraphics[width=\textwidth]{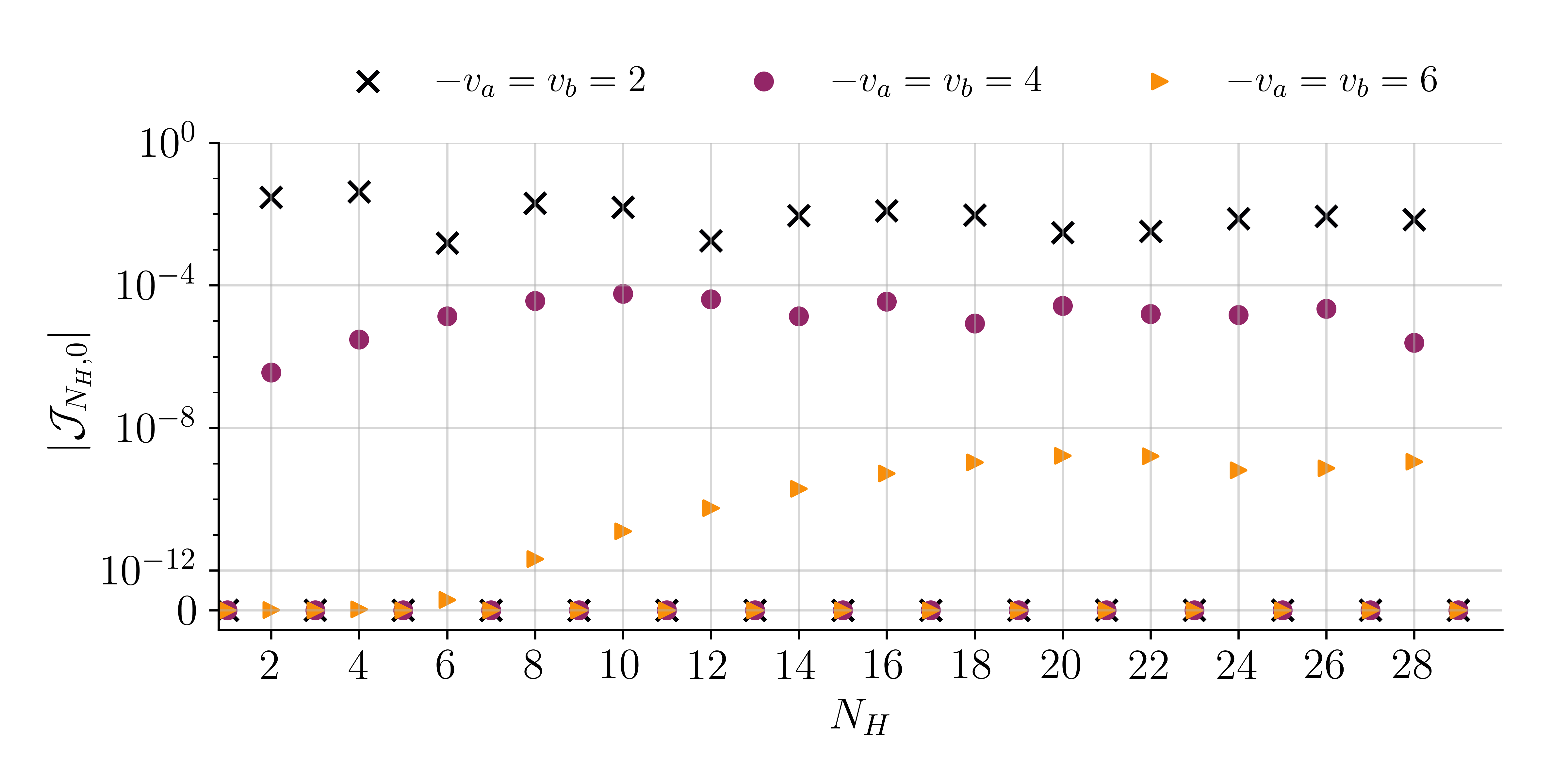}
    \end{subfigure}
    \hspace{-20pt}
    \begin{subfigure}[b]{0.51\textwidth}
        \centering 
        \caption{$|\mathcal{J}_{N_{H}, 1}|$}
        \includegraphics[width=\textwidth]{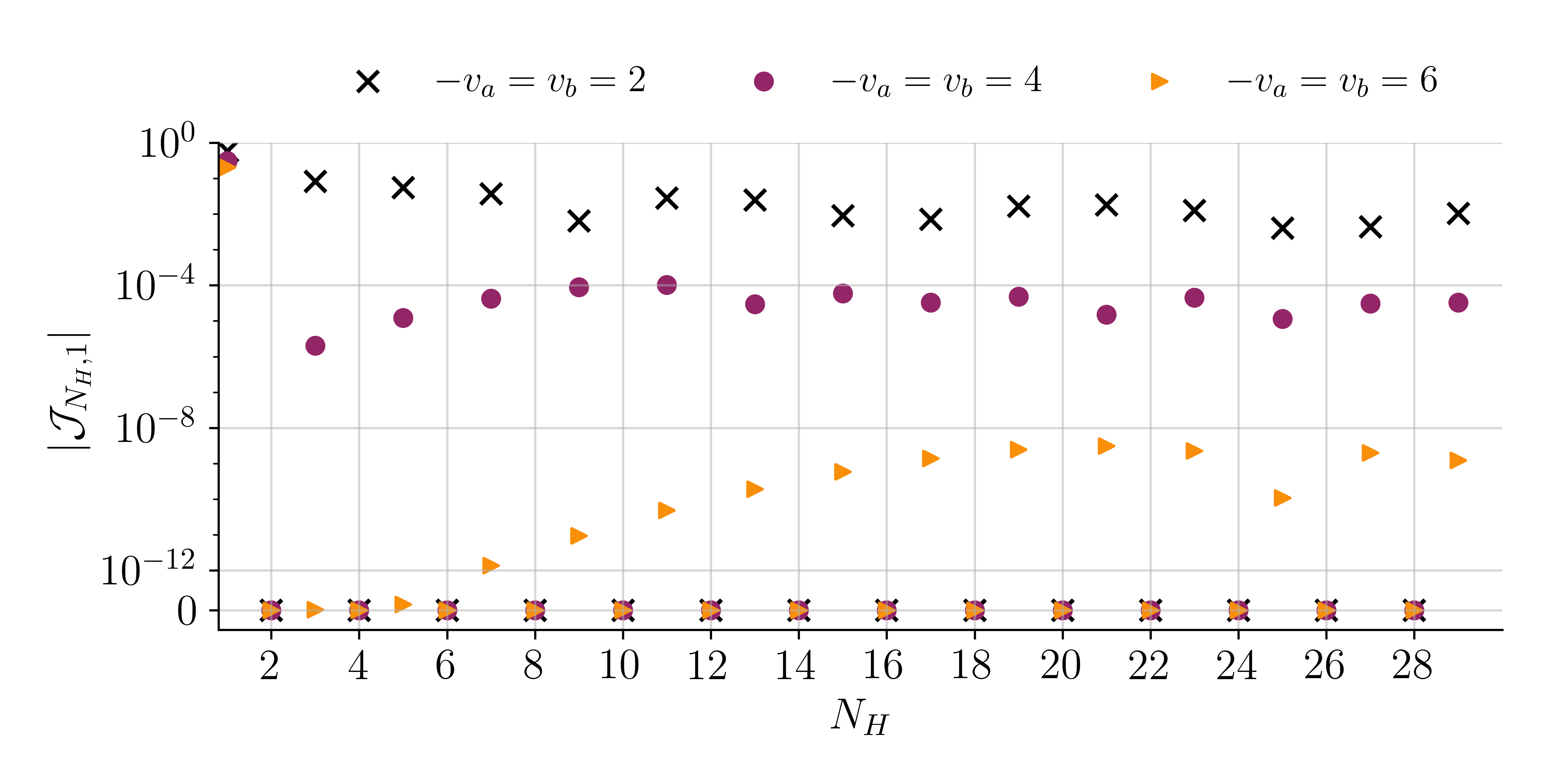}
    \end{subfigure}
    \begin{subfigure}[b]{0.51\textwidth}
        \centering 
        \caption{$|\mathcal{J}_{N_{H}, 2}|$}
        \includegraphics[width=\textwidth]{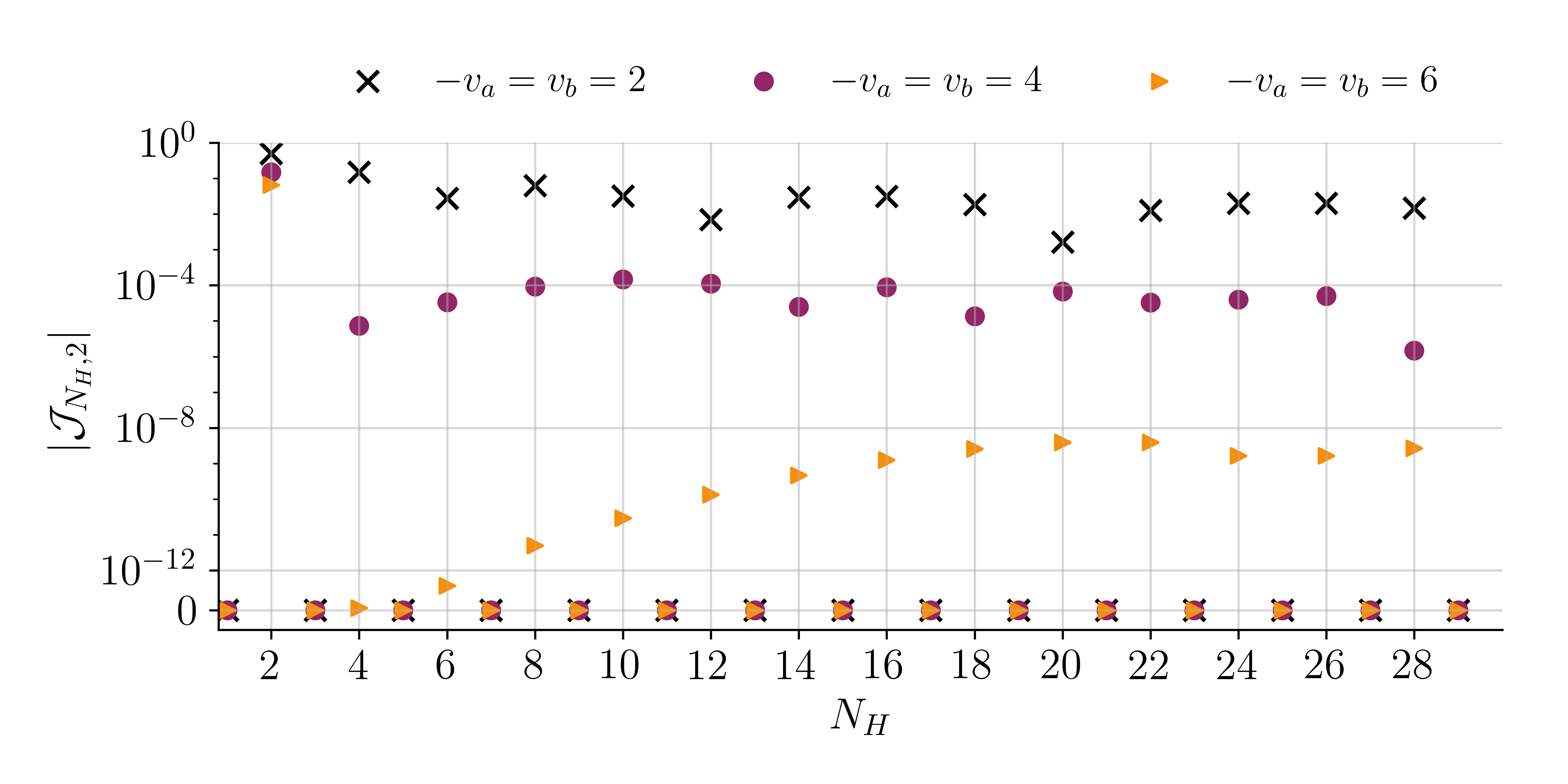}
    \end{subfigure}
    \caption{Numerical approximation of the integrals (a) $|\mathcal{J}_{N_{H}, 0}|$, (b) $|\mathcal{J}_{N_{H}, 1}|$, and (c) $|\mathcal{J}_{N_{H}, 2}|$ as a function of $N_{H}$ for various Legendre boundaries $-v_{a}= v_{b}$. The results show that as the Legendre domain increases, the integral values decrease. Additionally, the numerical results validate the analytic derivations in Eqns.~\eqref{J0}, \eqref{J1}, and~\eqref{J2}. }
    \label{fig:J_int}
\end{figure}

Let us summarize the conservation properties of the mixed method after discretization in velocity:
\begin{align}
    \frac{\mathrm{d}\mathcal{M}}{\mathrm{d} t} &= \frac{\sqrt{2N_{H}}}{\alpha} \mathcal{J}_{N_{H}, 0} \int_{0}^{\ell}  C_{N_{H}-1} \partial_{x} \phi\mathrm{d} x =  \begin{cases}
    0, \Hquad u = \bar{\sigma} \text{ and } N_{H} \text{ is \underline{odd}},\label{mass-conservation}\\
    0, \Hquad -v_{a} = v_{b} \to \infty \text{ and } N_{H} \geq 1, \\
    \text{non-zero}, \Hquad \text{else}.
    \end{cases} \\
    \frac{\mathrm{d}\mathcal{P}}{\mathrm{d} t} &= \frac{\sqrt{2N_{H}}}{\alpha} \left[\bar{\sigma} \mathcal{J}_{N_{H}, 0} + \sigma_{1}  \mathcal{J}_{N_{H}, 1}\right]\int_{0}^{\ell} C_{N_{H}-1} \partial_{x} \phi \mathrm{d} x = \begin{cases}
    0, \Hquad  -v_{a} = v_{b} \text{ and } u=0 \text{ and } N_{H} \text{ is \underline{even}},\\
    0, \Hquad -v_{a} = v_{b} \to \infty \text{ and } N_{H} \geq 2,\\
    \text{non-zero}, \Hquad \text{else}.
    \end{cases}\label{momentum-conservation}\\
    \frac{\mathrm{d}\mathcal{E}}{\mathrm{d} t} &=  \frac{\sqrt{2N_{H}}}{\alpha} \left[\frac{\sigma_{1}\sigma_{2}}{2} \mathcal{J}_{N_{H}, 2}+  \sigma_{1} \bar{\sigma} \mathcal{J}_{N_{H}, 1} + \frac{[\sigma_{1}^2 + \sigma_{0}^2 + \bar{\sigma}^2]}{2} \mathcal{J}_{N_{H}, 0} \right]\int_{0}^{\ell} C_{N_{H}-1} \partial_{x} \phi \mathrm{d} x \nonumber \\
    &= \begin{cases}
    0, \Hquad -v_{a} = v_{b} \text{ and } u=0 \text{ and } N_{H} \text{ is \underline{odd}}\\
    0, \Hquad -v_{a} = v_{b} \to \infty \text{ and } N_{H} \geq 3\\
    \text{non-zero}, \Hquad \text{else}.
    \end{cases}\label{energy-conservation}
\end{align}
Figure~\ref{fig:J_int} shows the integral values $\mathcal{J}_{N_{H}, 0}$ in Eq.~\eqref{J0}, $\mathcal{J}_{N_{H}, 1}$ in Eq.~\eqref{J1}, and $\mathcal{J}_{N_{H}, 2}$ in Eq.~\eqref{J2} as a function of the Hermite resolution $N_{H} \in \mathbb{N}_{>0}$ and for various Legendre boundaries $-v_{a} = v_{b}$.
The integral values are evaluated numerically using the trapezoidal rule with $N=10^{5}$ grid points in the velocity domain $v \in [v_{a}, v_{b}]$ and with the Hermite parameters $\alpha=1$ and $u=0$.
The numerical results validate the analytic results in Eqns.~\eqref{J0}, \eqref{J1}, and~\eqref{J2}, where  $\mathcal{J}_{N_{H}, 0} = \mathcal{J}_{N_{H}, 2} = 0$ if $N_{H}$ is odd and  $u = \bar{\sigma}$ and $\mathcal{J}_{N_{H}, 1} = 0$ if $N_{H}$ is even and  $u = \bar{\sigma}$. 
Additionally, as we increase the Legendre velocity bounds, the integral values decrease, which improves the conservation properties according to Eqns.~\eqref{mass-conservation}--\eqref{energy-conservation}.
However, as we increase the Legendre velocity bounds, we increase the smallest velocity scale resolved by $\delta f$, i.e., we lose accuracy.
Thus, in practice, we can guarantee either conservation of mass and energy or conservation of momentum alone, depending on $N_{H}$ parity.

To address this limitation, we propose a simple yet effective solution: we artificially enforce
\begin{equation*}
    \mathcal{J}_{N_{H}, 0}= \mathcal{J}_{N_{H}, 1} = \mathcal{J}_{N_{H}, 2}= 0,
\end{equation*}
thereby removing the coupling between the last Hermite coefficient $C_{N_{H}-1}$ and the first three Legendre coefficients $B_{0}$, $B_{1}$, and $B_{2}$. The impact of this approximation is assessed in section~\ref{sec:numerical_results}.

\section{Numerical results}\label{sec:numerical_results}
In this section, we test the mixed method on standard electrostatic benchmark problems.
The spatial discretization of Eqns.~\eqref{mixed_method_1_a_discrete}--\eqref{mixed_method_1_c_discrete}, temporal integrator, and parametric setup are described in section~\ref{sec:implementational_details}.
The numerical results for the linear advection problem are shown in section~\ref{sec:linear_advection}, the two-stream instability in section~\ref{sec:two_stream_instability}, and the bump-on-tail instability in section~\ref{sec:bump_on_tail_instability}.

\subsection{Implementation details}\label{sec:implementational_details}
We discretize the spatial periodic domain $x \in [0, \ell]$ with second-order central finite differencing. 
For the temporal integrator, we use a second-order implicit midpoint integrator since it conserves linear and quadratic invariants of the semi-discrete system~\cite{hairer_2006_sscm}, i.e., mass, momentum, and energy. 
The timestep is set to $\Delta t = 10^{-2}$ for all simulations. 
We solve the nonlinear system at each time step using an unpreconditioned Jacobian-Free-Newton-Krylov (JFNK)~\cite{knoll_2004_jcp} method with absolute and relative error tolerances set to $10^{-10}$. 
The linear Krylov solver that is embedded in the JFNK method is the default linear generalized minimum residual (LGMRES)~\cite{baker_2005_siam} method with relative and absolute tolerances set to $10^{-5}$. 
We solve the semi-discrete Poisson equation using the LGMRES method at each time step, with relative and absolute tolerances set to $10^{-12}$. 
All simulations in this paper were performed on a MacBook Pro 2.3 GHz Quad-Core Intel Core i7 processor with 16 GB RAM.
Moreover, in all simulations, we set the spatial grid resolution to $N_{x} = 101$ and the Legendre boundary condition penalty coefficient to $\gamma = 0.5$, consistent with~\citet[\S 3]{manzini_2016_jcp}.
The reference solution used to evaluate the accuracy of the mixed method for the two-stream instability in section~\ref{sec:two_stream_instability} and the bump-on-tail instability in section~\ref{sec:bump_on_tail_instability}, is a fully central finite difference simulation in space and velocity with Dirichlet boundary conditions in velocity with $f(x, v_{a}, t) = f(x, v_{b}, t) = 0$ as described in~\citet{shiroto_2019_fd} with $N_{x} =101$ and $N_{v} = 60{,}000$.

\subsection{Linear advection}\label{sec:linear_advection}
We begin with simulating the linear advection equation where $\phi(x, t) = 0$, i.e. $\partial_{t} f + v \partial_{x} f = 0$. The analytic solution of the advection equation is $f(x, v, t) = f(x-vt, v, t=0)$.
We initialize the electron distribution function as follows
\begin{equation*}
    f(x, v, t=0) = f_{0}(x, v, t=0) = \frac{1 + \cos(x)}{\sqrt{2\pi}}\exp\left(-\frac{v^2}{2}\right) \qquad \Rightarrow \qquad C_{0}(x, t=0) = \frac{1 + \cos(x)}{\sqrt{2}},
\end{equation*}
with $\delta f(x, v, t=0) = 0$, such that $B_{m}(x, t=0) = 0$ for $m=0, 1, \ldots, N_{L}-1$ and $C_{n}(x, t=0) = 0$ for $n=1, 2, \ldots, N_{H}-1$.
Moreover, we set the spatial domain length to $\ell = 2\pi$ and the artificial collisional frequencies to $\nu_{H} = \nu_{L} = 0$, so that no artificial collisions are included.
The Hermite scaling and shifting parameters are $\alpha = \sqrt{2}$ and $u = 0$, respectively. The Legendre velocity domain is defined by $v_{b} = -v_{a} = 5$, and the final simulation time is $T = 20$.

\begin{figure}
    \centering
    \includegraphics[width=\linewidth]{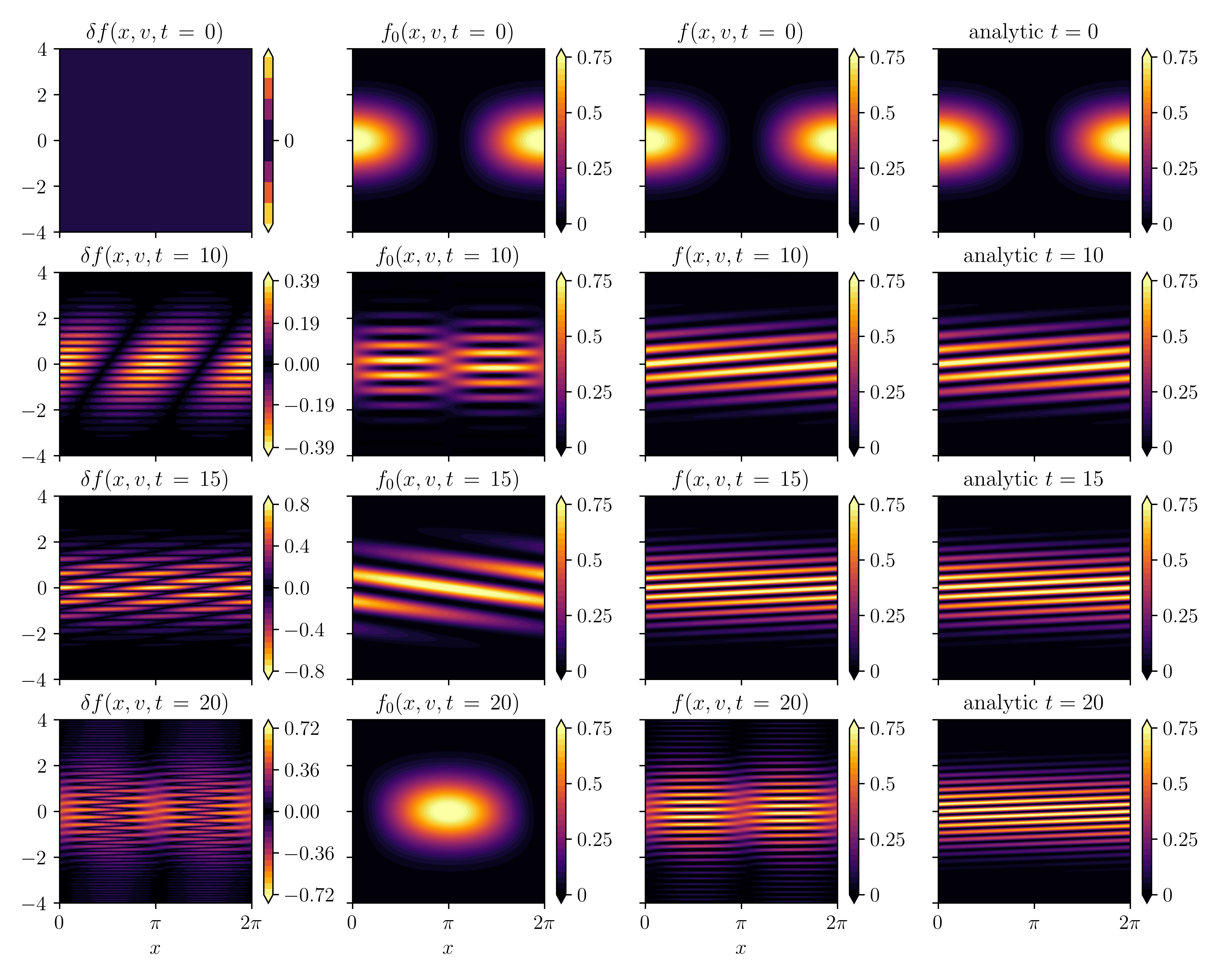}
    \caption{Comparison of the mixed method for the linear advection example with $N_{H}=N_{L}=100$ (third column) against the analytic solution $f(x, v, t) = f(x-vt, v, 0)$ (fourth column). Because the Hermite representation of $f_{0}(x, v, t)$  (second column) cannot resolve fine-scale filamentation early on, the Legendre representation of $\delta f(x, v, t)$ (first column) compensates by increasing in amplitude as time progresses (first column), until both discretizations run out of resolution in velocity space at around $t=20$ (fourth row).}
    \label{fig:advection_phase_space}
\end{figure}

\begin{figure}
    \centering
    \includegraphics[width=0.6\linewidth]{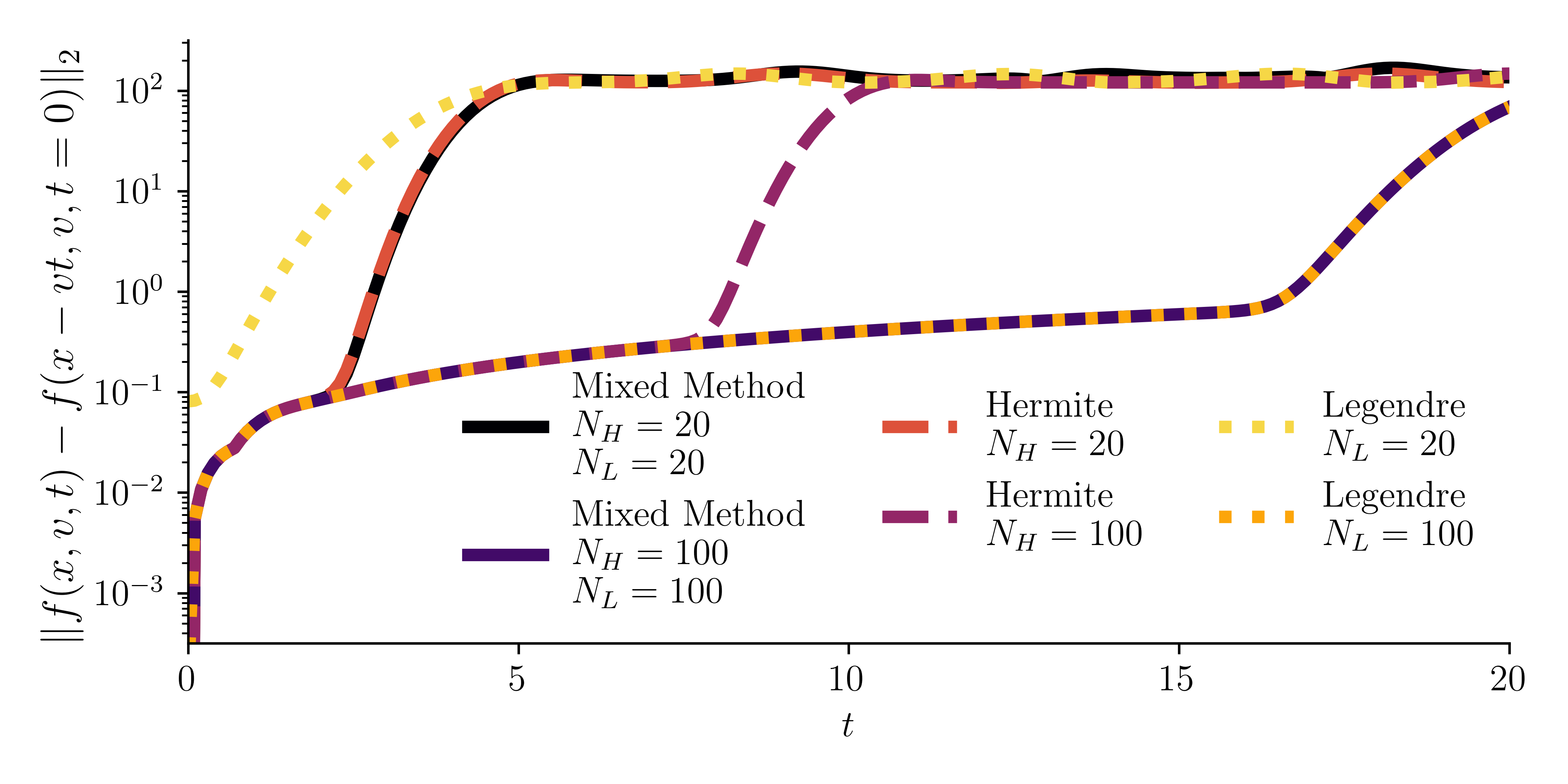}
    \caption{The electron distribution function $L_{2}$ error for the linear advection example using the mixed method and the individual Hermite and Legendre methods. The results show that the mixed method recurrence period is equivalent to the maximum of the individual Hermite or Legendre method recurrence period. }
    \label{fig:advection_error_analysis}
\end{figure}

Figure~\ref{fig:advection_phase_space} shows the evolution of the electron distribution function in phase space using the mixed method with $N_{H} = N_{L} = 100$.
At $t = 10$ (second row), the Hermite representation of $f_{0}$ runs out of velocity-space resolution and exhibits recurrence. The Hermite error is compensated by the Legendre representation of $\delta f$.
Thus, at $t = 10$ (second row) and $t = 15$ (third row), even though the Hermite description is inaccurate, the superposition of the Hermite and Legendre representation remains accurate when compared to the analytic solution shown in the last column.
By $t = 20$, the Legendre representation also runs out of velocity resolution, and the mixed method is no longer accurate.
To better understand the error and recurrence behavior of the mixed method, we compare its temporal evolution with that of the individual Hermite and Legendre methods in Figure~\ref{fig:advection_error_analysis}, using a consistent setup: $v_{b} = -v_{a} = 5$ for the Legendre method, and $u = 0$ with $\alpha = \sqrt{2}$ for the Hermite method.
The results indicate that the recurrence period of the mixed method matches the maximum recurrence period of the individual Hermite or Legendre method for the same number of respective spectral coefficients.
This agreement is expected, as the recurrence period is inversely proportional to the smallest velocity scale resolved by the spectral method. 
For a Hermite expansion with resolution $N_{H}$, this scale is $\lambda_{H} \propto \alpha/\sqrt{N_{H}}$, while for a Legendre expansion with resolution $N_{L}$, it is $\lambda_{L} \propto [v_{b}-v_{a}]/N_{L}$.
It is important to note that, in the present setup, the velocity bounds of the individual Legendre method and the Legendre component of the mixed method are identical, namely $v_{b} = -v_{a} = 5$. 
Consequently, no improvement in delaying the recurrence period is expected for the mixed method, as its performance remains limited by the smallest resolved velocity scale, which is $\min(\lambda_{L}, \lambda_{H})$.
For a Legendre expansion, this smallest velocity scale is linearly proportional to the size of the velocity domain, $v_{b} - v_{a}$. 
Therefore, when the domain is unchanged and the number of spectral terms remains the same, the attainable resolution, and hence recurrence period, remains unchanged as well.
Furthermore, Figure~\ref{fig:advection_error_analysis} shows two limiting cases. 
When $N_H=N_L=20$, the Hermite expansion is more accurate than the Legendre expansion and adding those Legendre terms in the mixed method does not delay recurrence. 
The case $N_H=N_L=100$ is opposite, as the Legendre expansion is now more accurate than the Hermite expansion. 
Hence, when the Hermite expansion loses accuracy the Legendre terms are able to compensate for it for some time (cf. Figure~\ref{fig:advection_phase_space}). 
In this case the Hermite terms are not helping with recurrence.
For a case like this, the results suggest that, instead of performing the simulation with a fixed number of Hermite and Legendre terms, it would be more advantageous to start the simulation with the Hermite expansion (since it captures near-Maxwellian behavior more efficiently), add Hermite terms until the accuracy of the Hermite expansion becomes comparable to that of the Legendre expansion with the same numbers of terms (recall the different asymptotic behavior of the error for the two expansions) and then switch completely to the Legendre expansion. 
As mentioned above, this behavior is due to the fact that in the mixed method we are applying the Legendre representation essentially over the whole velocity domain. Below we show with other examples how the mixed method can become more accurate than individual Hermite or Legendre representations when some strongly non-Maxwellian features are restricted to only a portion of phase space.
Overall, the mixed method behaves as intended, with the Legendre description compensating for the Hermite expansion and capturing the filamentary structures of the distribution function.

\begin{figure}
    \centering
    \begin{subfigure}[b]{0.45\textwidth}
        \centering 
        \caption{Hermite $|\hat{C}_{n}(k=1, t)|$}
        \includegraphics[width=\textwidth]{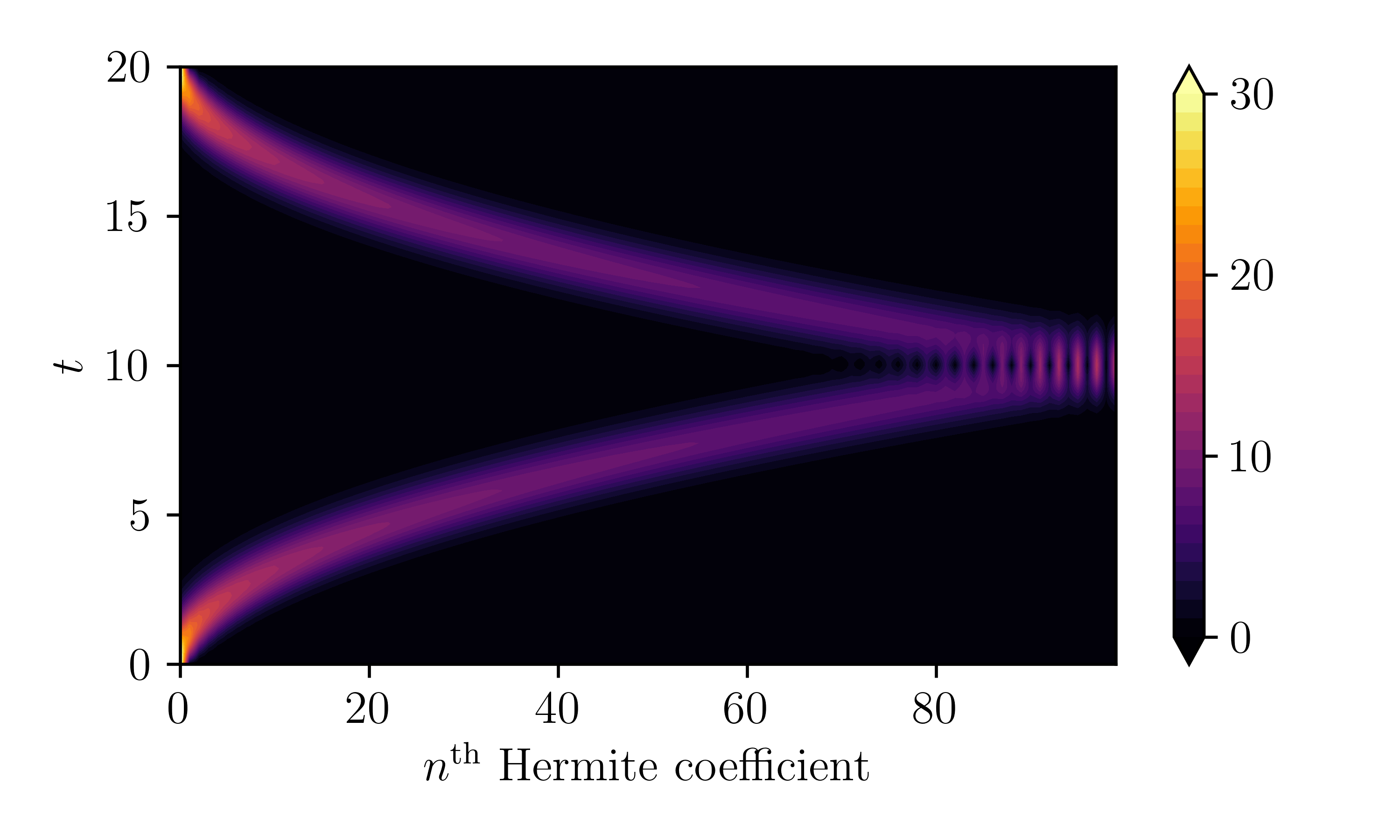}
    \end{subfigure}
    \begin{subfigure}[b]{0.45\textwidth}
        \centering 
        \caption{Legendre $|\hat{B}_{m}(k=1, t)|$}
        \includegraphics[width=\textwidth]{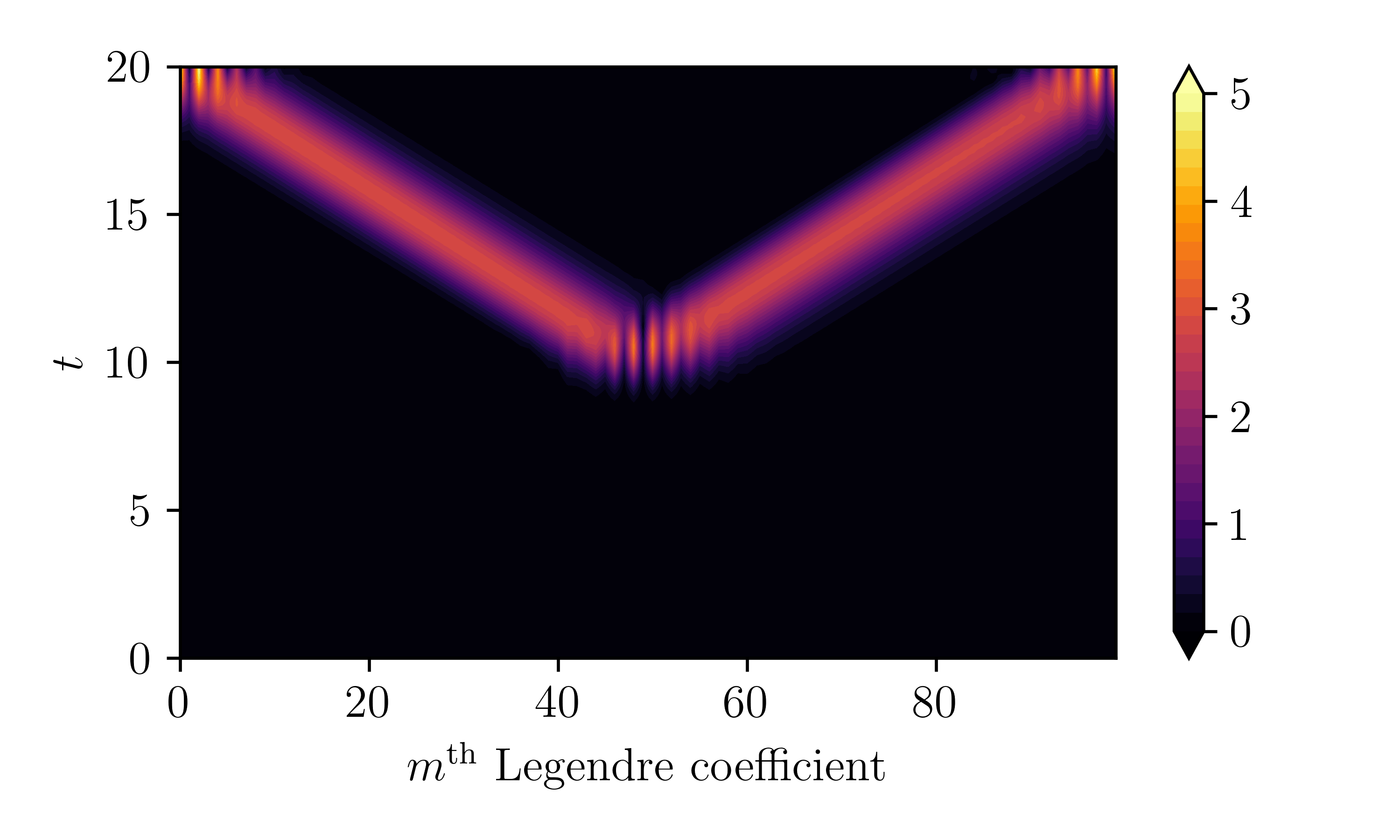}
    \end{subfigure}
    \caption{Linear advection temporal evolution of the (a) Hermite and (b) Legendre coefficients amplitudes at wavenumber $k=1$. The Hermite representation exhibits recurrence at $t \approx 10$, after which the Legendre amplitudes become more prominent and recur around $t \approx 18$.}
    \label{fig:advection-recurrence-fft-in-time}
\end{figure}

\begin{figure}
    \centering
    \begin{subfigure}[b]{0.45\textwidth}
        \centering 
        \caption{Hermite $|\hat{C}_{n}(k=1, t)|$}
        \includegraphics[width=\textwidth]{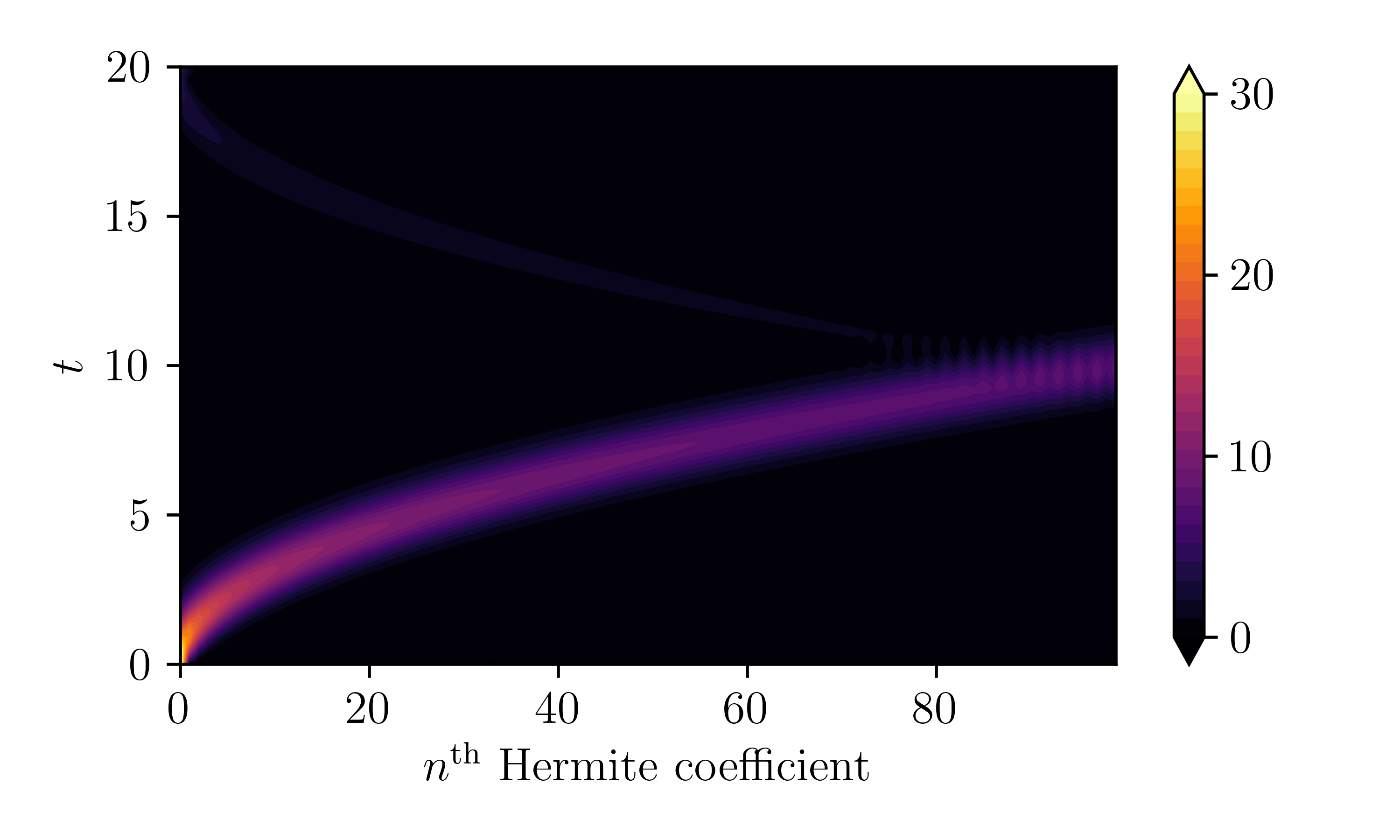}
    \end{subfigure}
    \begin{subfigure}[b]{0.45\textwidth}
        \centering 
        \caption{Legendre $|\hat{B}_{m}(k=1, t)|$}
        \includegraphics[width=\textwidth]{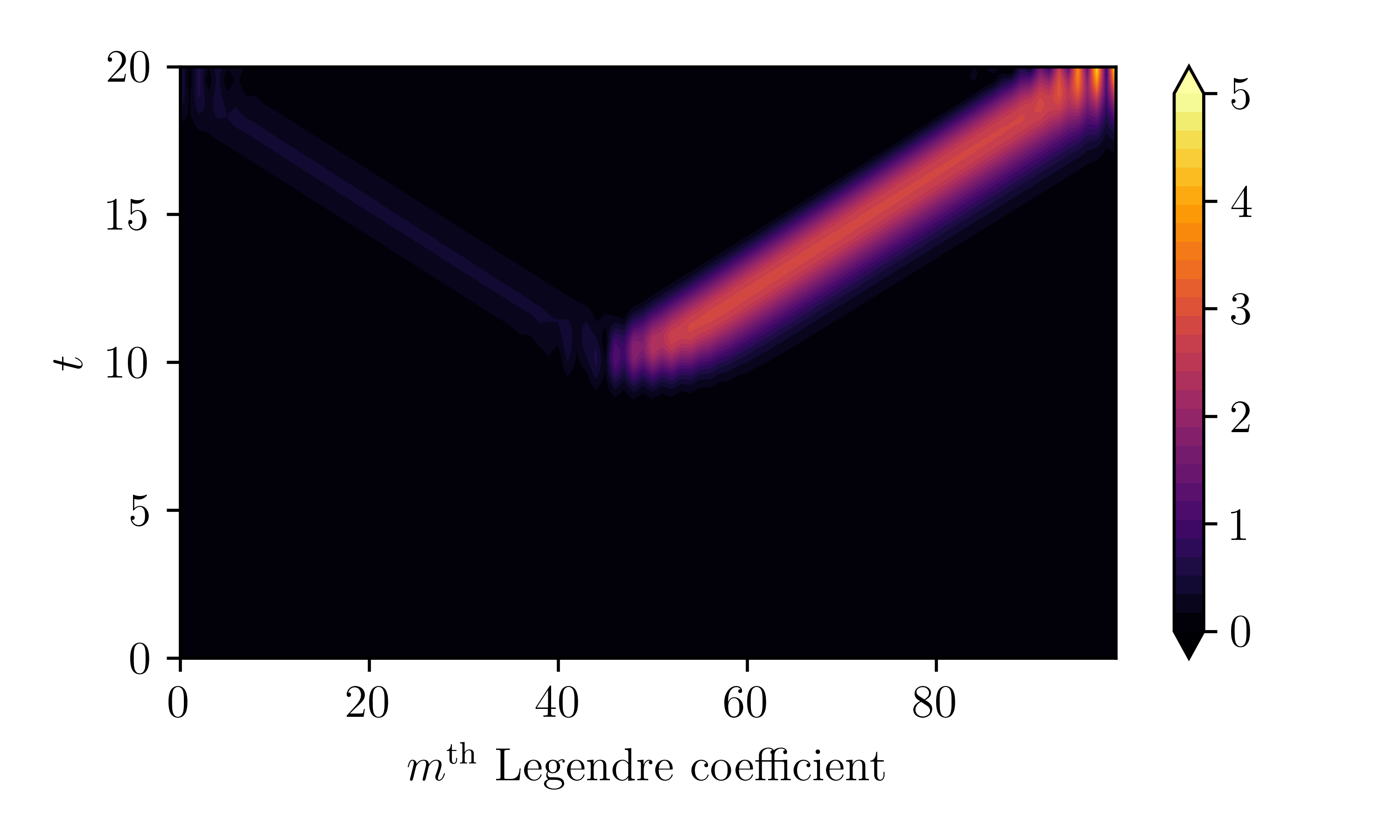}
    \end{subfigure}
    \caption{Same as Figure~\ref{fig:advection-recurrence-fft-in-time} with the closure in Eq.~\eqref{closure_df}, instead of closure by truncation ($C_{N_{H}} = 0$). The closure in Eq.~\eqref{closure_df} reduces the nonphysical backpropagation of Hermite flux, however it will still recur at the same time as the simulation with closure by truncation due to having the same finite spectral resolution $N_{H} = N_{L} = 100$.}
    \label{fig:advection-recurrence-fft-in-time-closure}
\end{figure}

\textit{A note on the closure.} Figure~\ref{fig:advection-recurrence-fft-in-time} shows the Hermite and Legendre coefficient amplitude cascade at wavenumber $k=1$ computed via the Fourier transform, for the case $N_H=N_L=100$ corresponding to Figure~\ref{fig:advection_phase_space}.
The results illustrate the transfer of energy from the zeroth-order Hermite coefficient to the highest-order Hermite coefficient, which is then reflected back to lower-order Hermite coefficients due to the finite Hermite resolution and the adopted closure. 
For the linear advection problem, the Hermite flux should propagate only from lower to higher-order Hermite coefficients, as the velocity space develops increasingly finer scales.
The excitation of the highest-order Hermite coefficient impacts the Legendre expansion, where information also eventually back-propagates at $t \approx 18$ due to finite Legendre resolution.
The nonphysical backpropagation of the Hermite coefficient flux shown in Figure~\ref{fig:advection-recurrence-fft-in-time} can be reduced by altering the closure, for example with nonlocal Landau closure~\cite{smith_1997_closure, issan_2024_pop} or if we set
\begin{equation}\label{closure_df}
    C_{N_{H}} =  \int_{v_{a}}^{v_{b}} \delta f \psi^{N_{H}} \mathrm{d} v = \sum_{m=0}^{N_{L}-1} B_{m} \mathcal{I}_{N_{H}, m}, \qquad \mathrm{where} \qquad \mathcal{I}_{n, m} \coloneqq \int_{v_{a}}^{v_{b}} \psi^{n}\xi_{m}\mathrm{d}v.
\end{equation}
Figure~\ref{fig:advection-recurrence-fft-in-time-closure} shows the Hermite and Legendre coefficients amplitude cascade at wavenumber $k=1$ computed via the Fourier transform, similar to Figure~\ref{fig:advection-recurrence-fft-in-time}. 
The only difference between the two figures is that Figure~\ref{fig:advection-recurrence-fft-in-time-closure} uses the closure in Eq.~\eqref{closure_df} and  Figure~\ref{fig:advection-recurrence-fft-in-time} uses \textit{closure by truncation}, i.e. $C_{N_{H}} = 0$.
The closure choice can reduce the nonphysical backpropagation of Hermite flux, but the simulation will still recur at the same time as the simulation with closure by truncation since the recurrence period is tied to the smallest velocity scales the method can resolve, which depends on the finite spectral resolution and parameters. 

\subsection{Two stream instability}\label{sec:two_stream_instability}
Next, we simulate the two-stream instability. 
We initialize the electron distribution function as follows
\begin{align*}
    f(x, v, t=0) &= f_{0}(x, v, t=0)
    = \frac{1+0.01\cos(0.5 x)}{\sqrt{2 \pi}} v^2 \exp\left(-\frac{v^2}{2}\right), \\
    C_{0}(x, t=0) &= \frac{1+0.01\cos\left(0.5 x\right)}{\sqrt{2}}, \\
    C_{2}(x, t=0) &= \sqrt{2}\, C_{0}(x, t=0),
\end{align*}
with $\delta f(x, v, t=0) = 0$, such that $B_{m}(x, t=0) = 0$ for $m = 0, 1, \ldots, N_{L}-1$.
In the Hermite expansion, only the modes $n=0$ and $n=2$ are nonzero at $t=0$, so that $C_{n}(x, t=0) = 0$ for $n=1$ and $n=3, 4, \ldots, N_{H}-1$.
Additionally, we set the spatial domain length to $\ell = 4\pi$, with $\nu_{L} = 1$ and $\nu_{H} = 0$, so that artificial collisions are applied only to the Legendre modes. The Hermite scaling and shifting parameters are $\alpha = \sqrt{2}$ and $u = 0$, and the Legendre velocity domain is defined by $v_{b} = -v_{a} = 2.5$. The final simulation time is $T = 35$, which is well into the nonlinear phase of the dynamics.

\begin{figure}
    \centering
    \begin{subfigure}[b]{0.49\textwidth}
        \centering 
        \caption{$N_{H} = 101$ with $\mathcal{J}_{ N_{H}, 1} \neq 0$}
        \includegraphics[width=\textwidth]{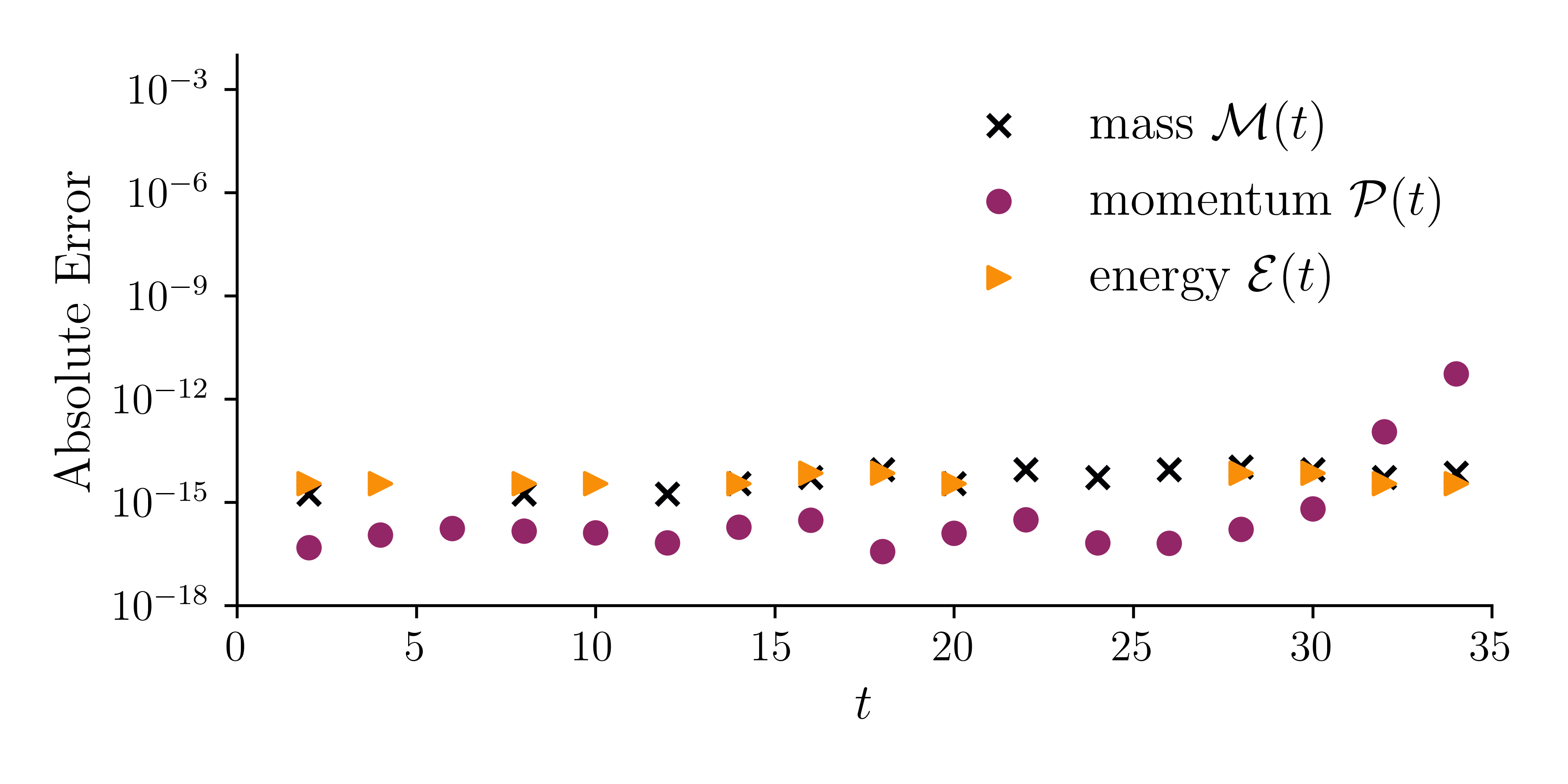}
        \label{fig:conservation-a}
    \end{subfigure}
    \begin{subfigure}[b]{0.49\textwidth}
        \centering 
        \caption{$N_{H}= 100$ with $\mathcal{J}_{ N_{H}, 0} \neq 0$ and $\mathcal{J}_{N_{H}, 2} \neq 0$}
        \includegraphics[width=\textwidth]{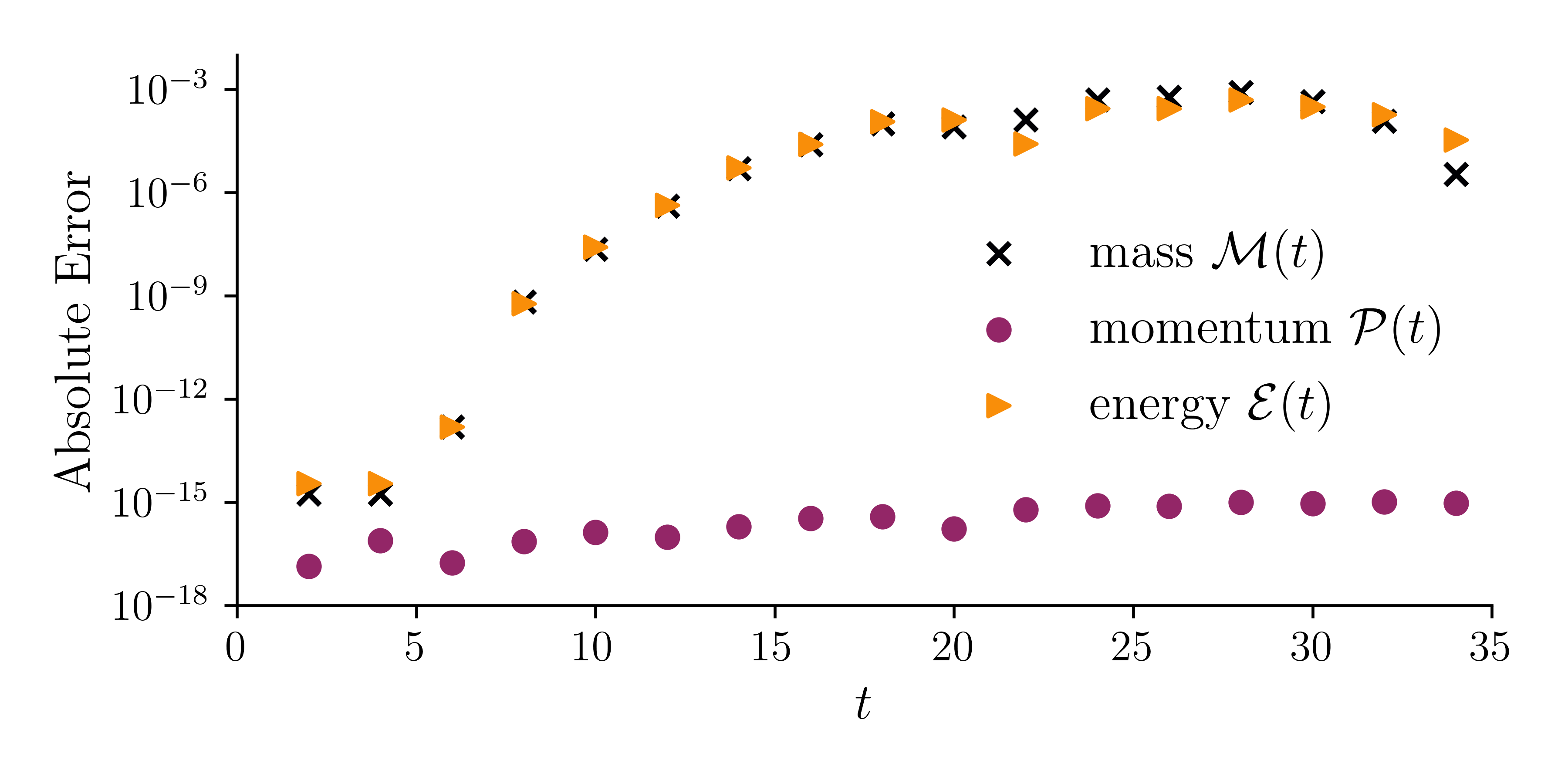}
         \label{fig:conservation-b}
    \end{subfigure}
    \begin{subfigure}[b]{0.49\textwidth}
        \centering 
        \caption{$N_{H} = 101$ with $\mathcal{J}_{ N_{H}, 1} = 0$}
        \includegraphics[width=\textwidth]{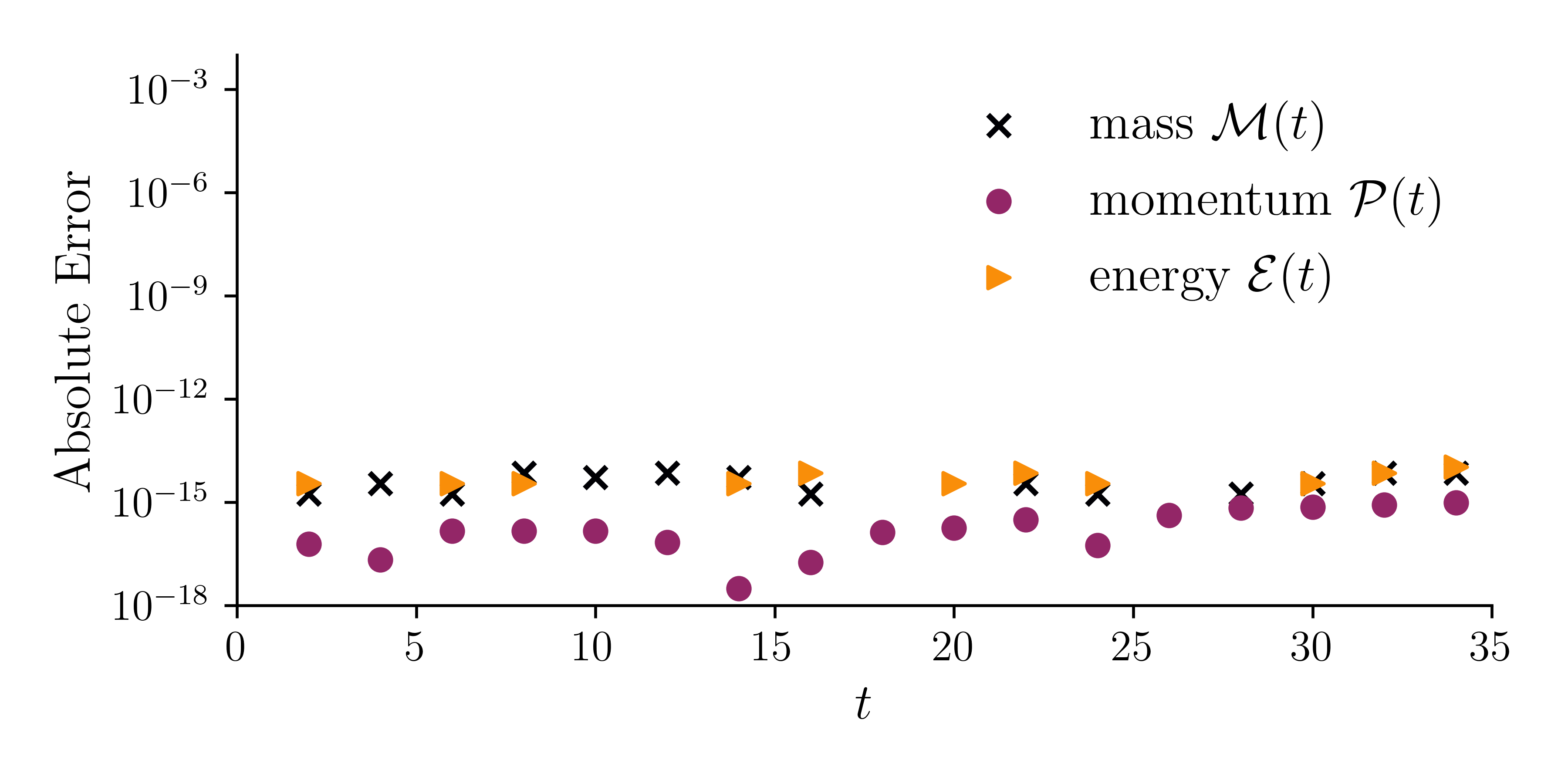}
        \label{fig:conservation-c}
    \end{subfigure}
    \begin{subfigure}[b]{0.49\textwidth}
        \centering 
        \caption{$N_{H} = 100$ with $\mathcal{J}_{ N_{H}, 0} =\mathcal{J}_{N_{H}, 2} = 0$}
        \includegraphics[width=\textwidth]{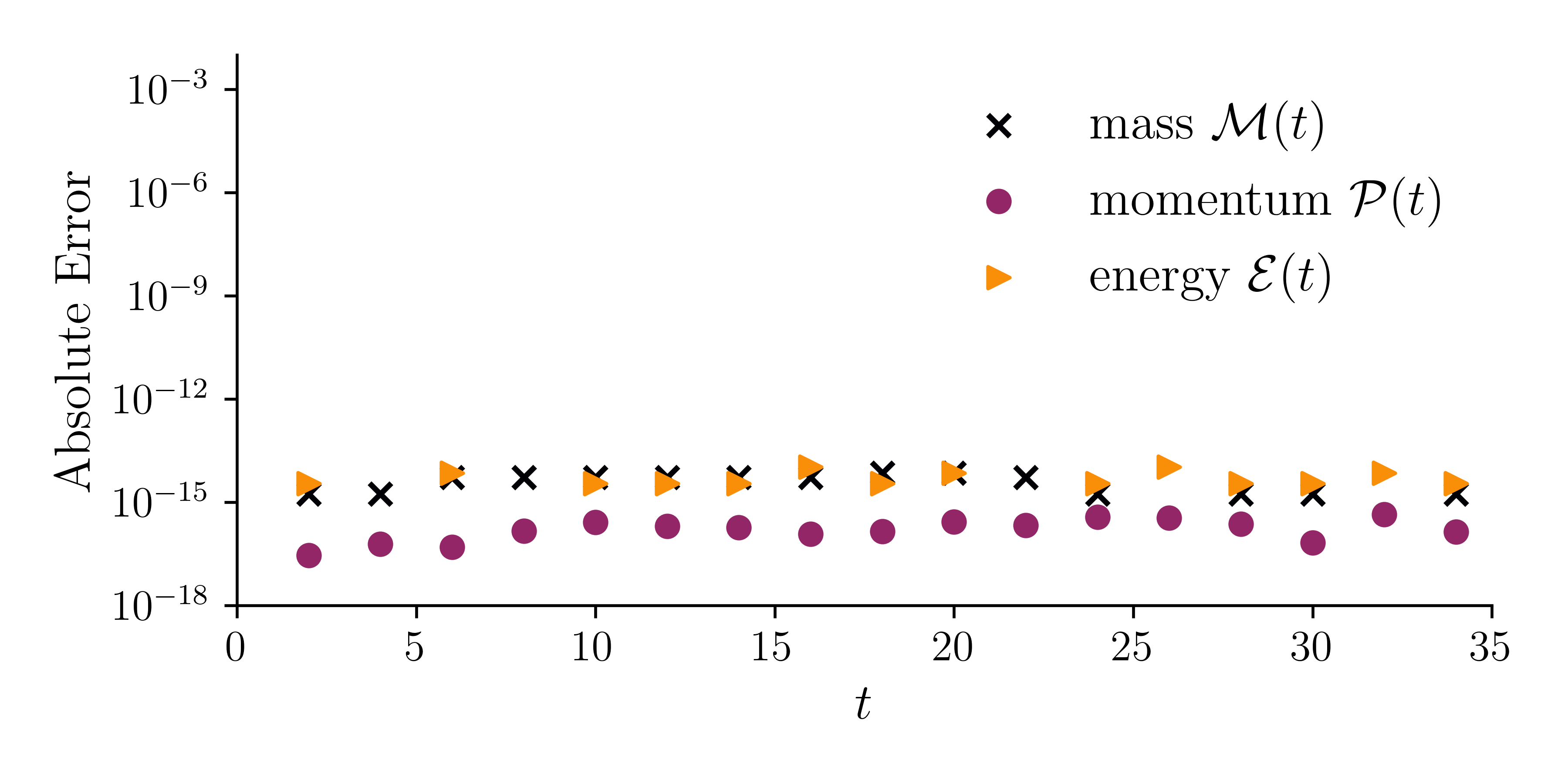}
        \label{fig:conservation-d}
    \end{subfigure}
    \caption{Two-stream instability: mass, momentum, and energy conservation for (a/c) $N_{H} = 101$ and (b/d) $N_{H} = 100$. All simulations above have $N_{L} = 100$. Subfigures (a/b) show results without enforcing $\mathcal{J}_{N_{H}, 0} = \mathcal{J}_{N_{H}, 1} = \mathcal{J}_{N_{H}, 2} = 0$, while (c/d) include this enforcement. The conservation behavior agrees with the analytic predictions of section~\ref{sec:conservation_laws}, where mass and energy are conserved for $-v_{a} = v_{b}$ when $N_{H}$ is odd (subfigure a), and momentum is conserved when $-v_{a} = v_{b}$ and $N_{H}$ is even (subfigure b). With the proposed enforcement $\mathcal{J}_{N_{H}, 0} = \mathcal{J}_{N_{H}, 1} = \mathcal{J}_{N_{H}, 2} = 0$, all conservation laws are satisfied (subfigures c and d). }
    \label{fig:two_stream_conservation_laws}
\end{figure}

Figure~\ref{fig:two_stream_conservation_laws} shows the mixed-method numerical conservation of mass, momentum, and energy, which confirms the analytical derivations presented in section~\ref{sec:conservation_laws}.
Without enforcing $\mathcal{J}_{N_{H},0} = \mathcal{J}_{N_{H},1} = \mathcal{J}_{N_{H},2} = 0$ (subfigures~\ref{fig:conservation-a} and \ref{fig:conservation-b}), the discrete conservation properties depend solely on $N_{H}$ when the velocity domain is symmetric, i.e., $-v_{a} = v_{b}$.
For odd $N_{H}$, for example $N_{H} = 101$ in subfigure~\ref{fig:conservation-a}, only mass and energy are conserved.
Conversely, for even $N_{H}$, for example $N_{H} = 100$ in subfigure~\ref{fig:conservation-b}, only momentum is conserved.
When $\mathcal{J}_{N_{H},0} = \mathcal{J}_{N_{H},1} = \mathcal{J}_{N_{H},2} = 0$ is enforced, as shown in subfigures~\ref{fig:conservation-c} and~\ref{fig:conservation-d}, all three invariants, mass, momentum, and energy, are simultaneously conserved (up to the nonlinear solver tolerance of $10^{-10}$), independently of the parity of $N_{H}$. 
The relative error in the electron distribution function between simulations with and without the enforced constraint is very small. For $N_{H} = 100$, the relative error is $0.06\%$, while for $N_{H} = 101$ it is $0.28\%$.
The $L_{2}$ relative error is computed by normalizing the difference with respect to the simulation without the constraint.
Therefore, enforcing the constraint does not significantly alter the simulation results while preserving the conservation properties of the system. For the remainder of this subsection, we therefore present results obtained with the constraint $\mathcal{J}_{N_H,0} = \mathcal{J}_{N_H,1} = \mathcal{J}_{N_H,2} = 0$ enforced.

\begin{figure}
    \centering
    \includegraphics[width=\linewidth]{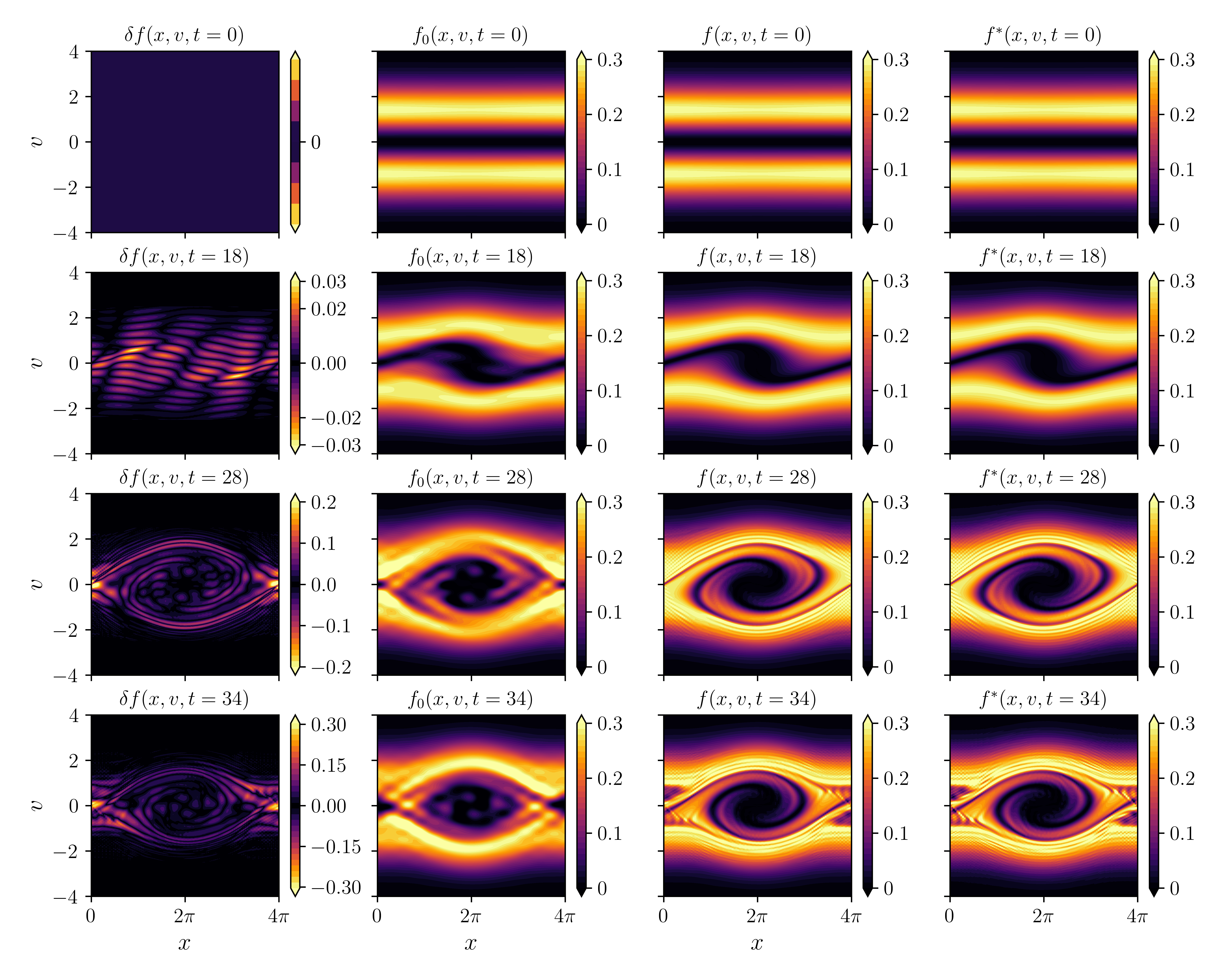}
    \caption{Two-stream instability electron distribution function in phase space with $N_{H}=85$ and $N_{L}=171$ ($N_{H} + N_{L} = 256$). The last column shows a reference solution $f^{*}(x, v, t)$ computed using a fully second-order central finite difference solver with resolution $N_{v}=60{,}000$. As the distribution develops highly non-Maxwellian features, $\delta f(x, v, t)$ grows in magnitude and becomes comparable to $f_{0}(x, v, t)$ around $t \approx 28$ (third row), yet their superposition remains accurate when compared to the reference solution. }
    \label{fig:two_stream_in_phase_space}
\end{figure}

Figure~\ref{fig:two_stream_in_phase_space} shows the electron distribution function obtained with the mixed method for the two-stream instability. 
The first column shows the Legendre representation of $\delta f$ with $N_{L} = 171$.
The second column shows the Hermite representation of $f_{0}$ with $N_{H} = 85$. 
The third column shows their superposition, $f = f_{0} + \delta f$, with a total of $N_{H}+ N_{L} = 256$ DOFs in velocity. 
The fourth column shows the reference solution obtained with a fully finite-difference solver using $N_{v} = 60{,}000$ velocity grid points.
The mixed-method results are in good agreement with the reference solution. 
As expected, the Hermite representation of $f_{0}$ does not accurately capture the emerging non-Maxwellian features, i.e., the phase space vortex dynamics, which are instead resolved by the Legendre component of the mixed method.

\begin{figure}
    \centering
    \begin{subfigure}[b]{0.45\textwidth}
        \centering 
        \caption{Distribution function error}
        \includegraphics[width=\textwidth]{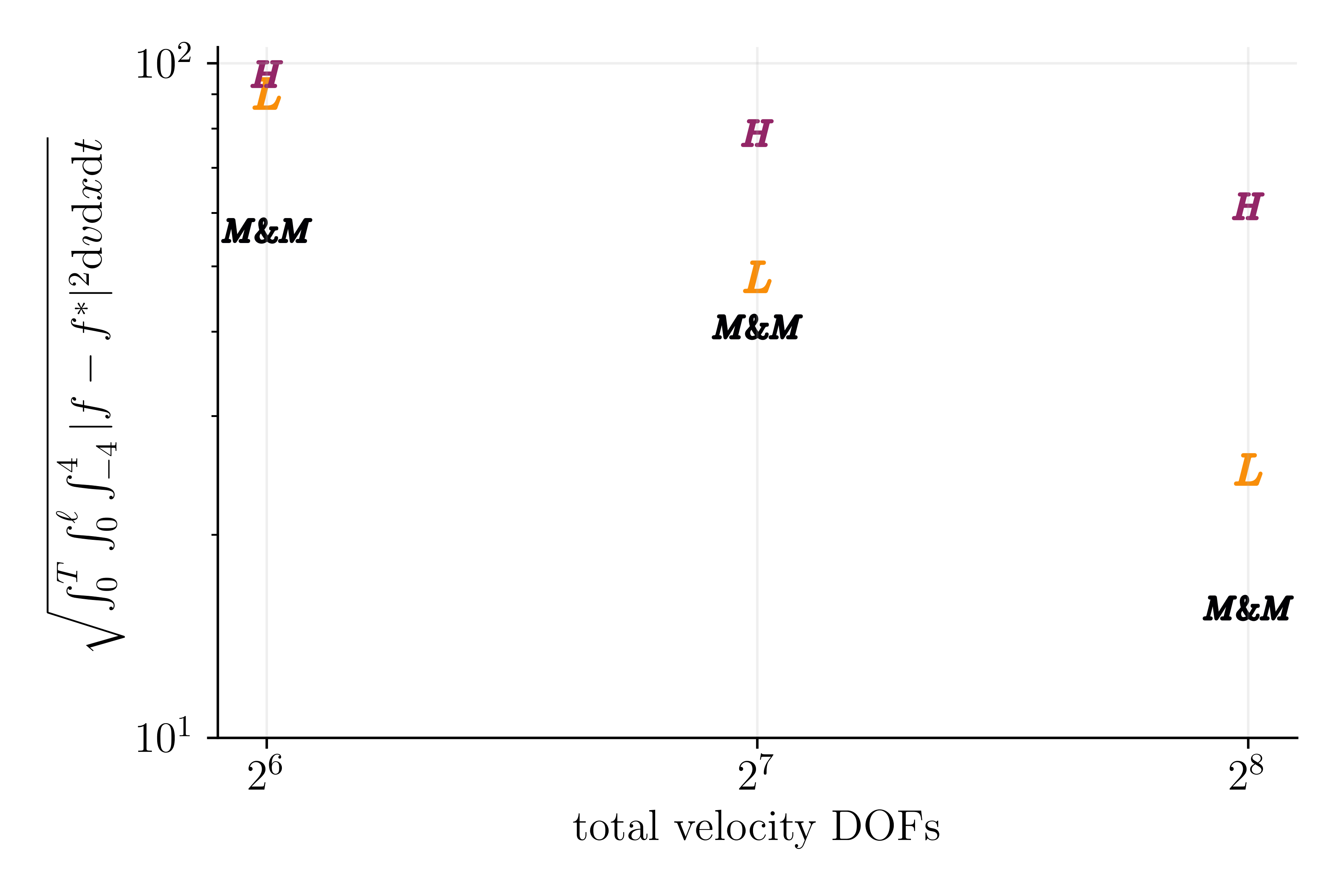}
    \end{subfigure}
    \hspace{-10pt}
    \begin{subfigure}[b]{0.45\textwidth}
        \centering 
        \caption{CPU runtime}
        \includegraphics[width=\textwidth]{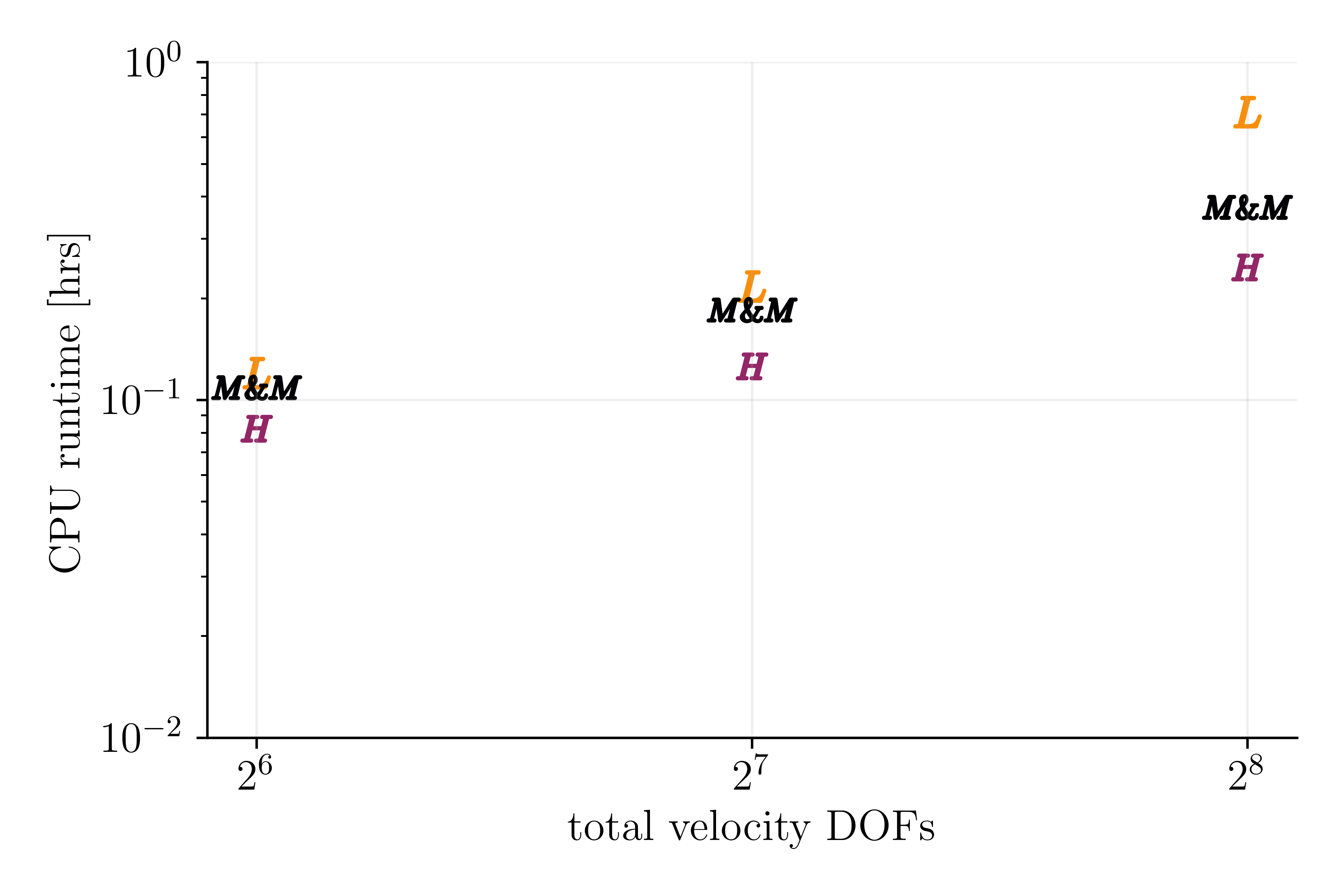}
    \end{subfigure}
    \caption{Two-stream instability (a) distribution function $L_{2}$ error and (b) CPU runtime comparison between the mixed method (labeled $\mathrm{M\& M}$ with $N_{H} = \lfloor N_{L}/2 \rfloor$), and the individual Hermite and Legendre methods (labeled $\mathrm{H}$ and $\mathrm{L}$ respectively). The errors are evaluated relative to the reference finite difference solution. In all the mixed method simulations, we set $\mathcal{J}_{N_{H}, 0} = \mathcal{J}_{N_{H}, 1} = \mathcal{J}_{N_{H}, 2} = 0$.}
    \label{fig:mixed_method_error_runtime_two_stream}
\end{figure}

Figure~\ref{fig:mixed_method_error_runtime_two_stream} shows the mixed method and the individual Hermite/Legendre methods distribution function error and CPU runtime as functions of the total number of velocity DOFs. 
We refer to the `individual Hermite/Legendre methods as the spectral method with only Hermite or just Legendre expansion of $f$ in velocity. The latter with $v_{b} = -v_{a} = 5$.
The error is computed against the reference finite difference solution.
Although we have not made any attempt to optimize the performance of each individual code, the computational effort is broadly comparable across methods: the mixed method is slightly less demanding than the Legendre method, yet slightly more demanding than the Hermite method.
The mixed method consistently achieves lower error than the individual Hermite and individual Legendre methods with the same total DOFs in velocity (and space).
The two-stream instability is most difficult for the individual Hermite method since the distribution develops strong non-Maxwellian features, which the Legendre method can accurately capture. 
The mixed method restricts the Legendre bounds to $v_{a} = -v_{b} = 2.5$, which is smaller than the individual Legendre method with $v_{a} = -v_{b} = 5$, such that restricting the Legendre bounds is key to the effectiveness of the mixed method. This example also illustrates one of the advantages of the mixed method, i.e. its flexibility relative to an approach that treats the distribution function with only Hermite or Legendre basis functions over the whole velocity space domain.

\subsection{Bump-on-tail instability}\label{sec:bump_on_tail_instability}
For the final test case, we consider the bump-on-tail instability with parameters matching those in~\citet{chapurin_2024_pop}. 
The initial electron distribution function is decomposed into a bulk Maxwellian and a beam-shifted Maxwellian:
\begin{align*}
    f_{0}(x, v, 0) &= \frac{1 - n_{\mathrm{beam}}}{\sqrt{2\pi}}
    \left[1 + \epsilon \cos\!\left(\frac{x}{10}\right)\right]
    \exp\!\left(-\frac{v^{2}}{2}\right), \\
    \delta f(x, v, 0) &= \frac{n_{\mathrm{beam}}}{\sqrt{2\pi}}
    \exp\!\left(-\frac{(v - 10)^{2}}{2}\right), \\
    C_{0}(x, 0) &= \frac{1 - n_{\mathrm{beam}}}{\sqrt{2}}
    \left[1 + \epsilon \cos\!\left(\frac{x}{10}\right)\right],
\end{align*}
where $n_{\mathrm{beam}} = 0.01$, $\epsilon = 10^{-4}$, $\ell = 20\pi$, $v_{a} = 4$, $v_{b} = 15$, $\nu_{H} = 10$, $\nu_{L} = 1$, $\alpha = \sqrt{2}$, $u = 0$, and final time $T=120$. 
The bulk Maxwellian is represented exactly by the zeroth Hermite coefficient, such that $C_{n}(x, 0) = 0$ for $n = 1, 2, \ldots, N_{H}-1$. 
The mixed method Legendre coefficients are computed by numerically projecting the beam $\delta f(x, v, t= 0)$ onto the Legendre basis, and the required integrals are evaluated using the trapezoidal rule with $N = 10^{4}$ quadrature points.
Since the non-Maxwellian features associated with the beam are well separated from the Maxwellian bulk, unlike in the two-stream instability, the coupling between the two components is weak. Therefore, for the mixed method, it is reasonable to set a large Hermite collision frequency $\nu_{H} = 10$.
In this section, we later compare the mixed method to the individual Hermite and Legendre methods, where for the individual Hermite method we separate the bulk and beam as two different electron distributions that satisfy the Vlasov equation, with $u_{e1} = 0$, $u_{e2} = 10$, and $\alpha_{e1} = \alpha_{e2} = \sqrt{2}$. 
Moreover, for the individual Legendre method, we set the velocity boundaries at $v_{a} = -5$ and $v_{b} = 15$. 

\begin{figure}
    \centering
    \begin{subfigure}[b]{0.49\textwidth}
        \centering 
        \caption{$N_{H} = 16$ with \newline $\mathcal{J}_{ N_{H}, 0} \neq 0$ and $\mathcal{J}_{ N_{H}, 0} \neq 1$ and $\mathcal{J}_{ N_{H}, 2} \neq 0$ }
        \includegraphics[width=\textwidth]{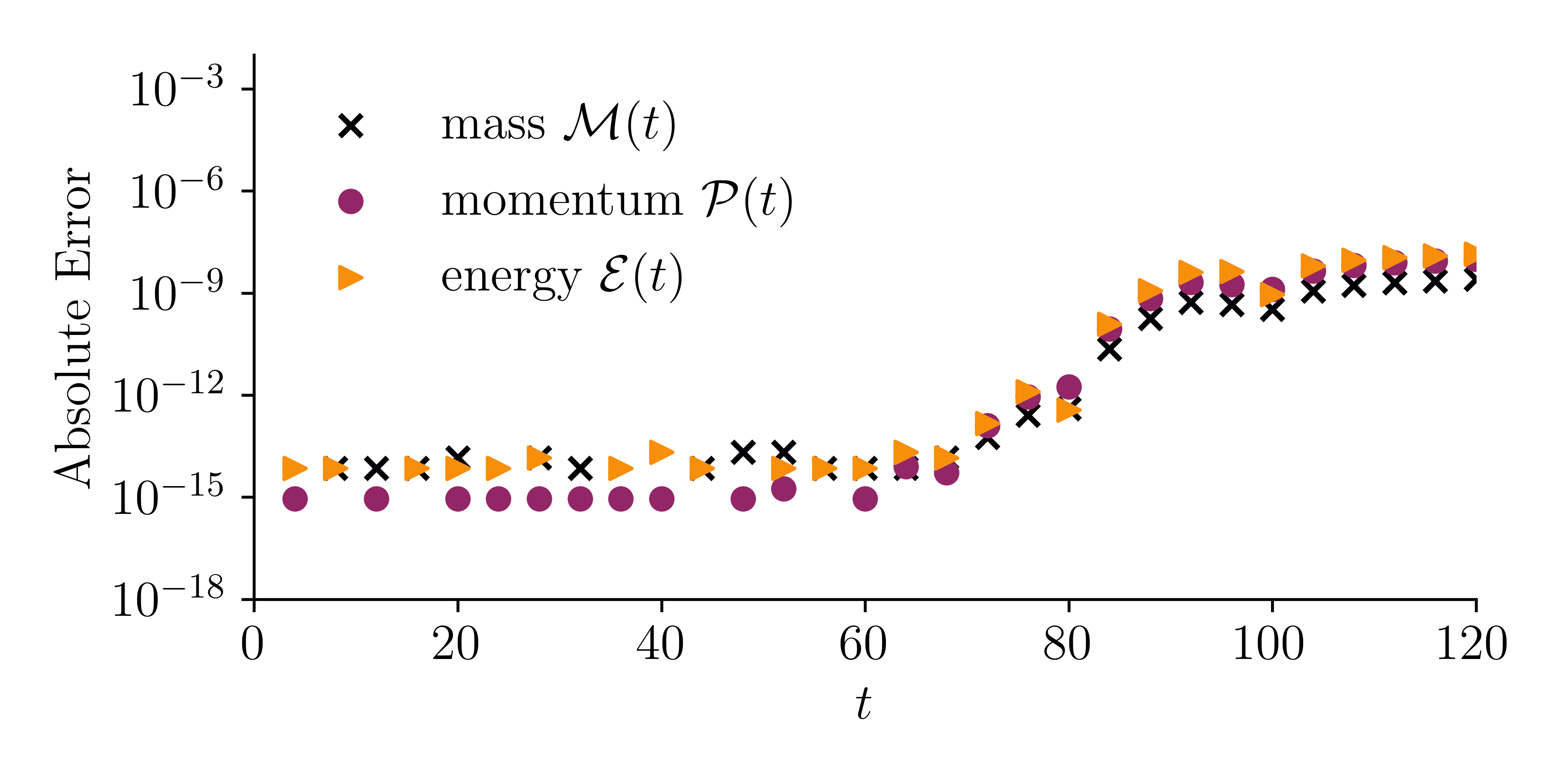}
        \label{fig:conservation-a-BOT}
    \end{subfigure}
    \begin{subfigure}[b]{0.49\textwidth}
        \centering 
        \caption{$N_{H} = 17$ with \newline $\mathcal{J}_{ N_{H}, 0} \neq 0$ and $\mathcal{J}_{ N_{H}, 0} \neq 1$ and $\mathcal{J}_{ N_{H}, 2} \neq 0$ }
        \includegraphics[width=\textwidth]{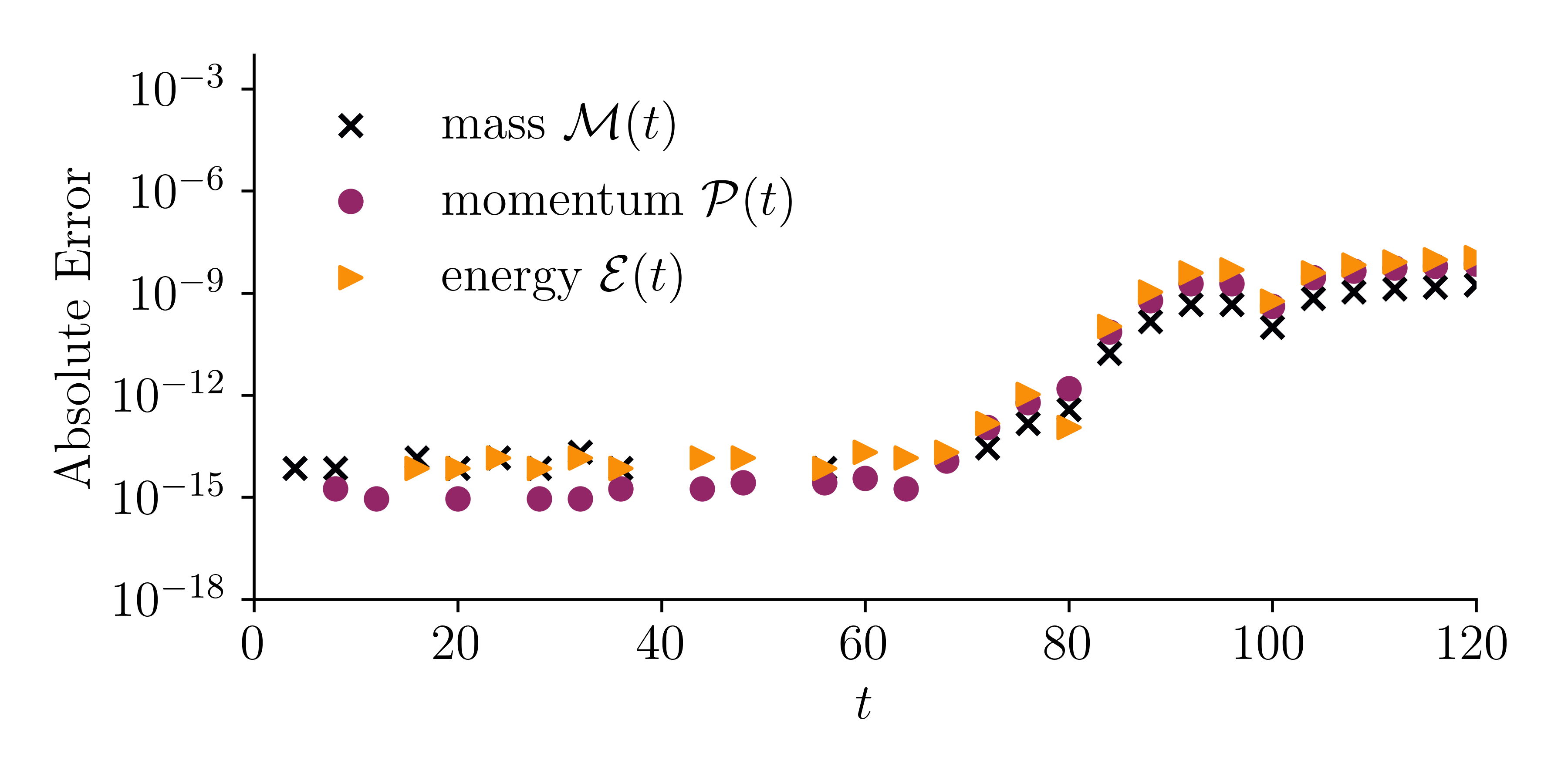}
         \label{fig:conservation-b-BOT}
    \end{subfigure}
    \begin{subfigure}[b]{0.49\textwidth}
        \centering 
        \caption{$N_{H} = 16$ with setting \newline $\mathcal{J}_{ N_{H}, 0} = \mathcal{J}_{ N_{H}, 1}=\mathcal{J}_{N_{H}, 2} = 0$}
        \includegraphics[width=\textwidth]{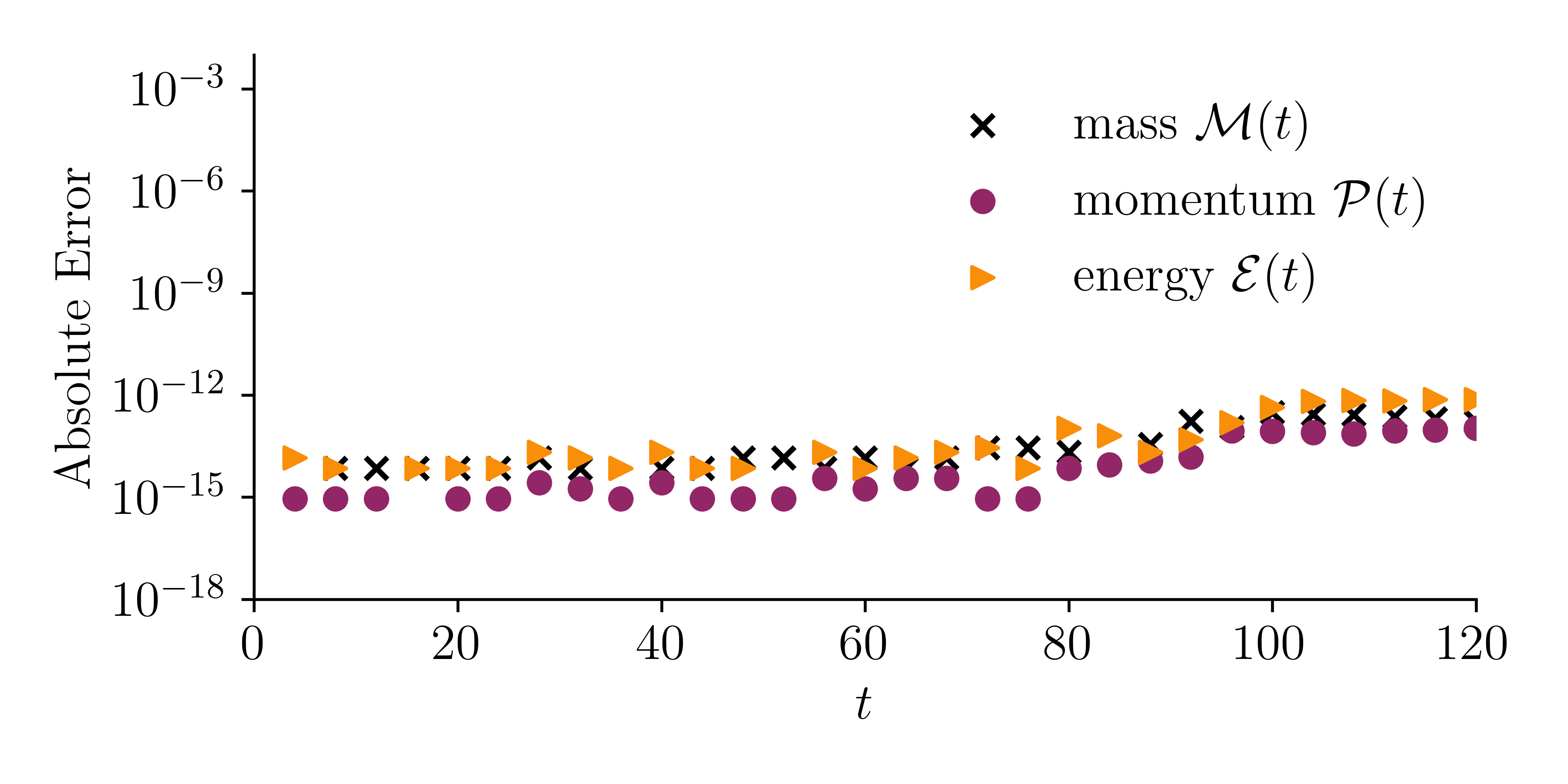}
        \label{fig:conservation-c-BOT}
    \end{subfigure}
    \begin{subfigure}[b]{0.49\textwidth}
        \centering 
        \caption{$N_{H} =17$ with setting \newline $\mathcal{J}_{ N_{H}, 0} = \mathcal{J}_{ N_{H}, 1}=\mathcal{J}_{N_{H}, 2} = 0$}
        \includegraphics[width=\textwidth]{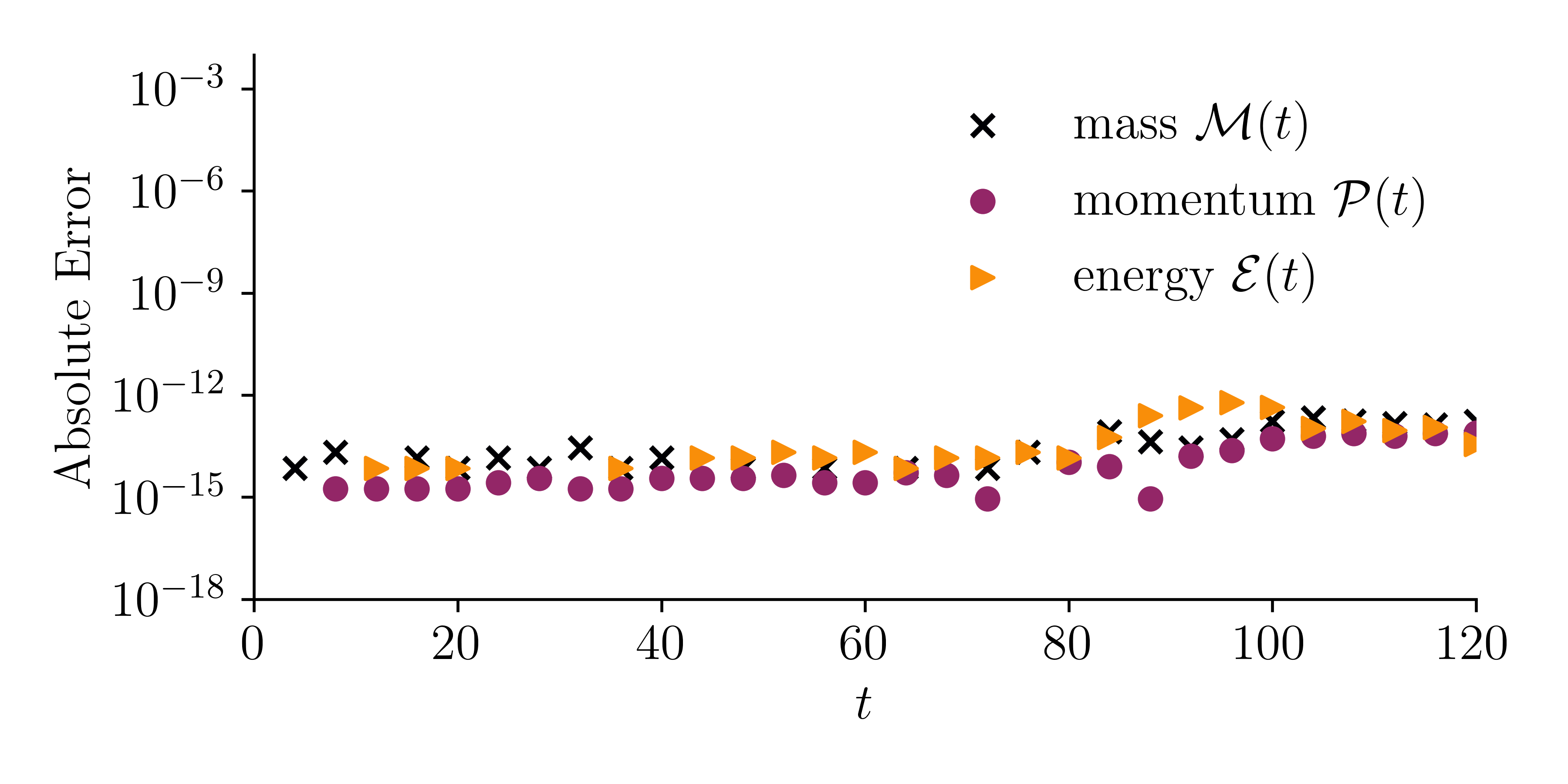}
        \label{fig:conservation-d-BOT}
    \end{subfigure}
    \caption{Same as Figure~\ref{fig:two_stream_conservation_laws} for the bump-on-tail instability. In all simulations $N_{L} = 64$ and the velocity bounds are $v_{a}=4$ and $v_{b}=15$, so that $|v_{a}| \neq |v_{b}|$. Consequently, for any choice of $N_{H}$ (odd or even), the integrals satisfy $\mathcal{J}_{N_{H},0} \neq 0$, $\mathcal{J}_{N_{H},1} \neq 0$, and $\mathcal{J}_{N_{H},2} \neq 0$, which violates mass, momentum, and energy conservation, as shown in subfigures~(a/b). In subfigures~(c/d), we set $\mathcal{J}_{N_{H},0} = \mathcal{J}_{N_{H},1} = \mathcal{J}_{N_{H},2} = 0$, which leads to conservation of mass, momentum, and energy up to the nonlinear solver tolerance.}
    \label{fig:bump_on_tail_conservation_laws}
\end{figure}

Since we set $v_{a} = 4$ and $v_{b} = 15$ for the mixed method, then it follows from section~\ref{sec:conservation_summary} that the conservation of mass, momentum, and energy are violated regardless of $N_{H}$ being even or odd since $|v_{a} | \neq |v_{b}|$.  
This is shown in subfigure~\ref{fig:conservation-a-BOT} with $N_{H} = 16$ and subfigure~\ref{fig:conservation-b-BOT} with $N_{H} = 17$.  
However, if we set $\mathcal{J}_{N_{H}, 0} =\mathcal{J}_{N_{H}, 1}=\mathcal{J}_{N_{H}, 2}=0$, then mass, momentum, and energy are conserved up to the temporal nonlinear solver tolerance (which we set to $10^{-10}$), as shown in subfigure~\ref{fig:conservation-c-BOT} with $N_{H} = 16$ and subfigure~\ref{fig:conservation-d-BOT} with $N_{H} = 17$.
The effect of enforcing the constraint on the electron distribution function is negligible.
The $L_{2}$ relative error between simulations with and without the constraint is $6 \times 10^{-8}\%$ for $N_{H} = 16$ and $3\times 10^{-8}\%$ for $N_{H} = 17$, where the relative error is computed by normalizing the difference with respect to the simulation without the constraint.
The relative error is very small for this example since the coupling terms are weak due to the convergence of the Hermite expansion (the bulk remaining close to Maxwellian). 
These results indicate that enforcing the constraint preserves the conservation properties of the system without significantly altering the simulation outcome. Therefore, in the remainder of this subsection we present results obtained with the constraint $\mathcal{J}_{N_H,0} = \mathcal{J}_{N_H,1} = \mathcal{J}_{N_H,2} = 0$ enforced.

\begin{figure}
    \centering
    \begin{subfigure}[b]{0.25\textwidth}
        \centering 
        \caption{Hermite \\ $N_{H_{1}} = 16$ and $N_{H_{2}} = 48$}
        \includegraphics[width=\textwidth]{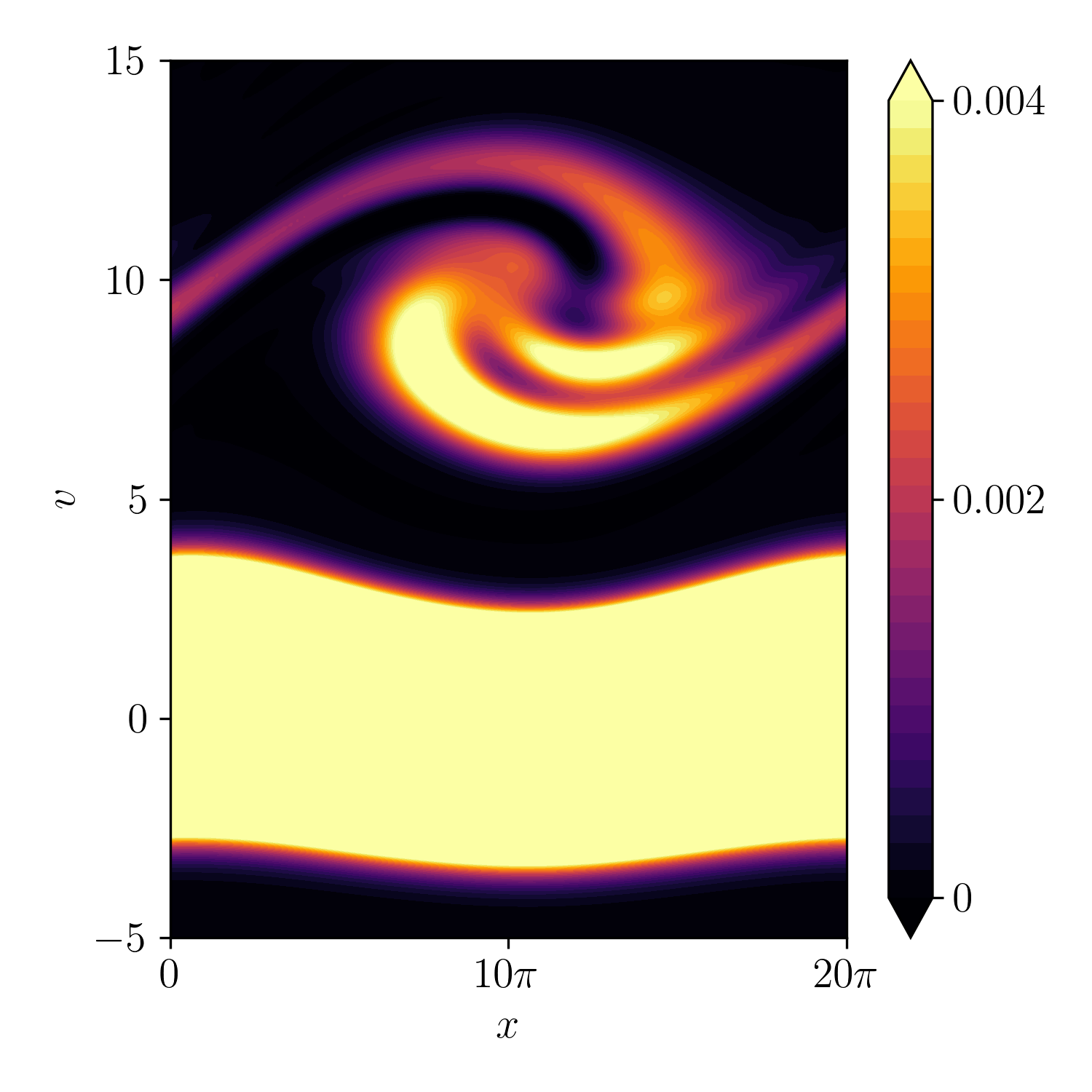}
    \end{subfigure}
    \begin{subfigure}[b]{0.25\textwidth}
        \centering 
        \caption{Legendre \\ $N_{v} = 64$}
        \includegraphics[width=\textwidth]{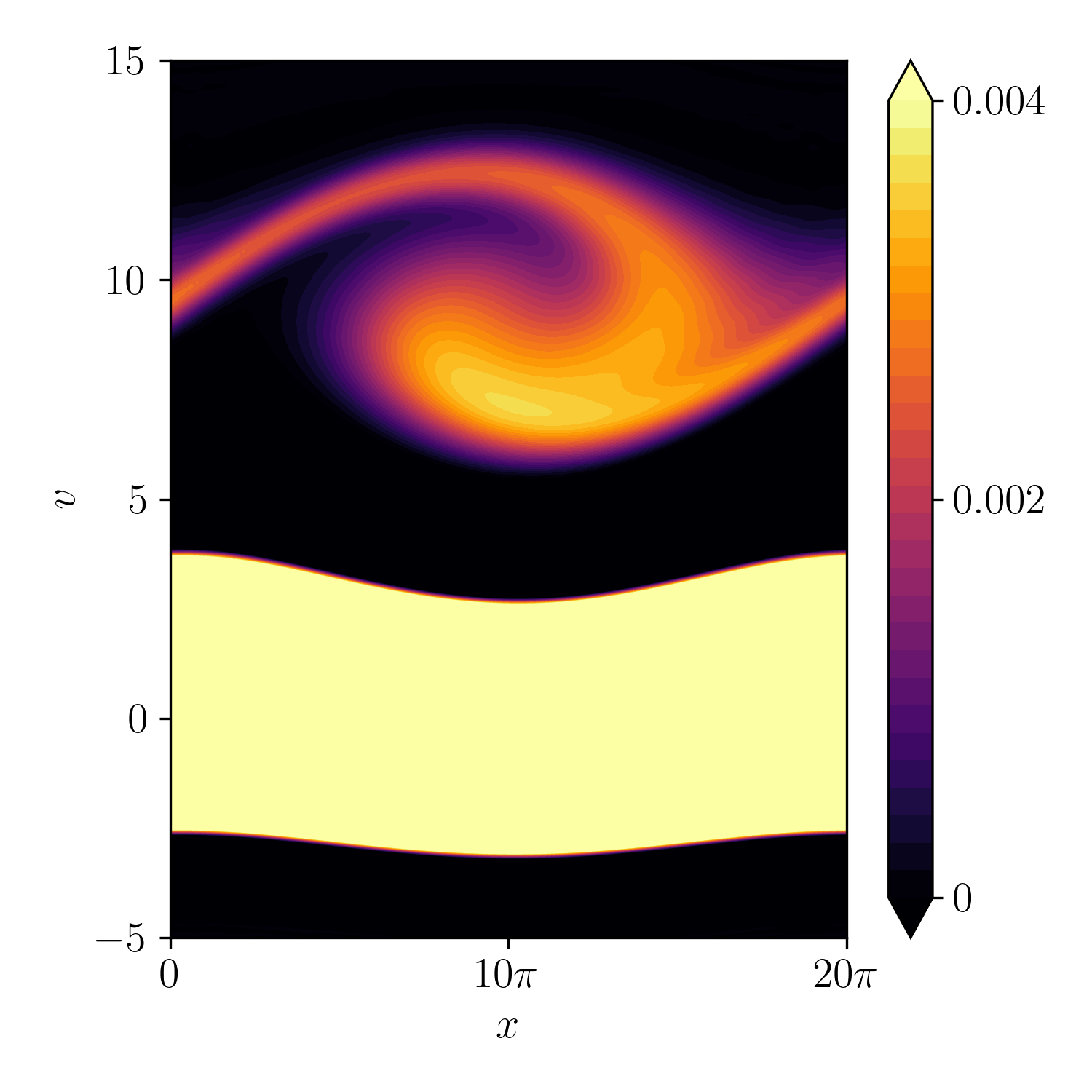}
    \end{subfigure}
    \begin{subfigure}[b]{0.25\textwidth}
        \centering 
        \caption{Mixed method \\$N_{H} = 16$ and $N_{L} = 48$}
        \includegraphics[width=\textwidth]{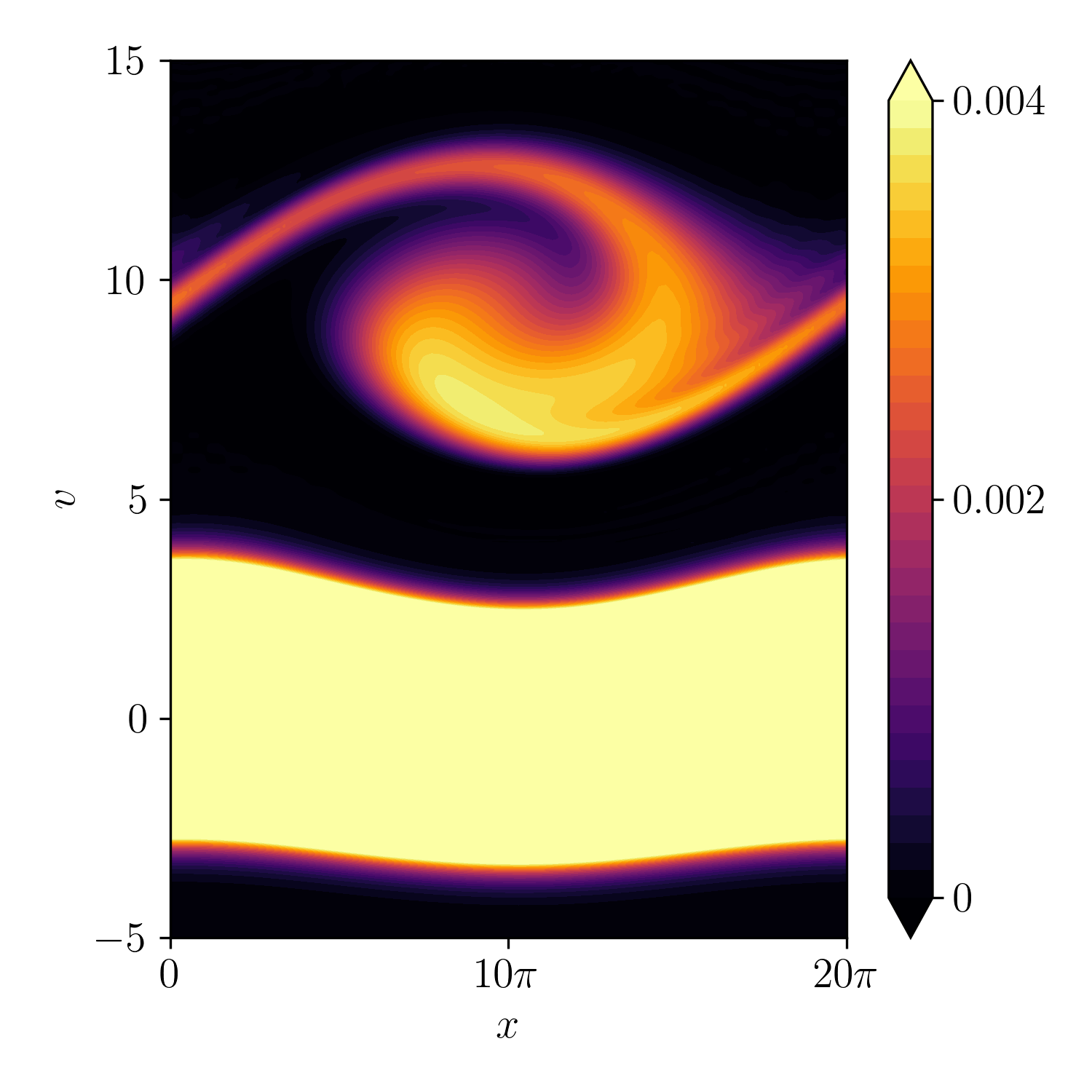}
    \end{subfigure}
        \begin{subfigure}[b]{0.25\textwidth}
        \centering 
        \caption{Mixed method \\$N_{H} = 16$ and $N_{L} = 112$}
        \includegraphics[width=\textwidth]{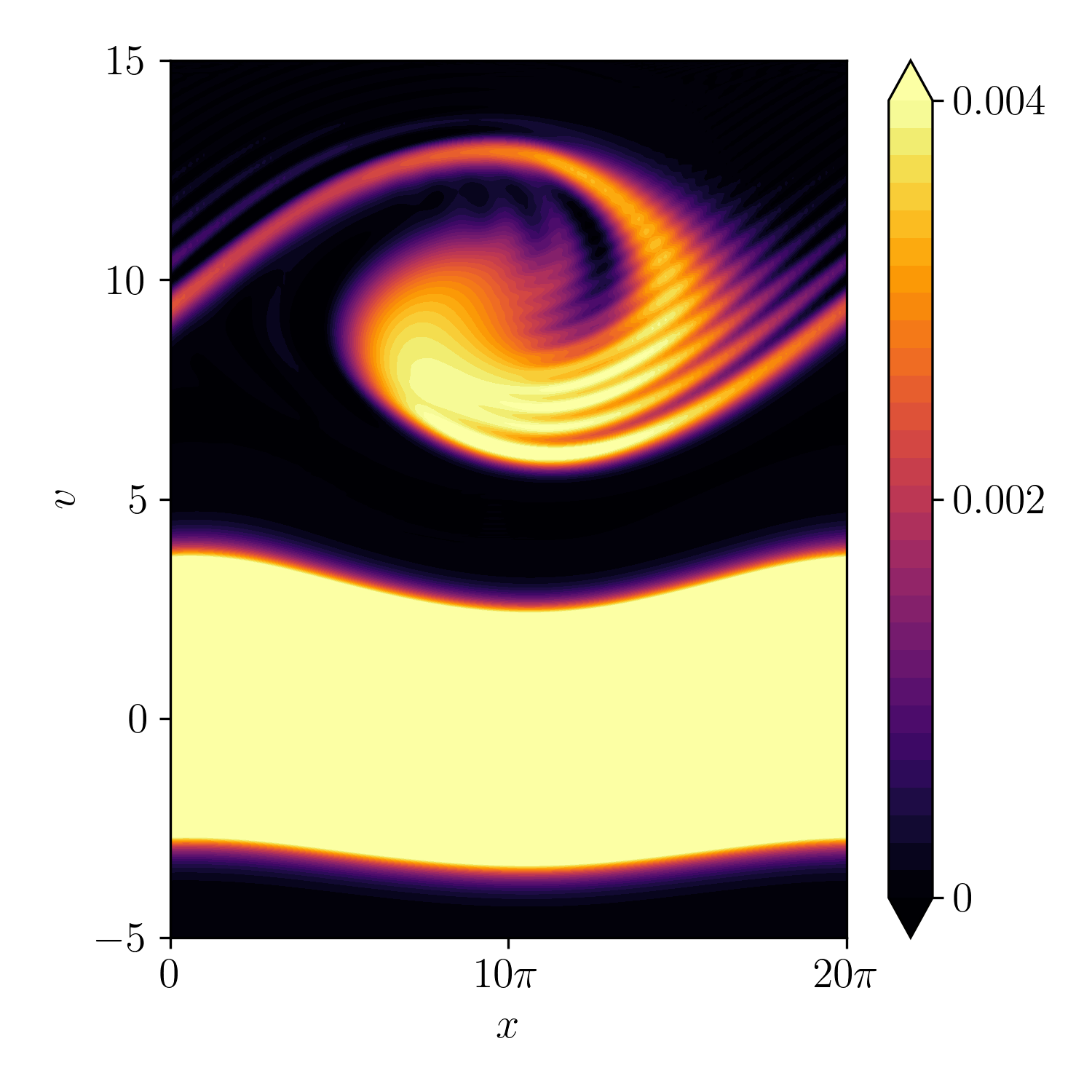}
    \end{subfigure}
        \begin{subfigure}[b]{0.25\textwidth}
        \centering 
        \caption{Reference (finite difference) \\$N_{v} = 60,000$}
        \includegraphics[width=\textwidth]{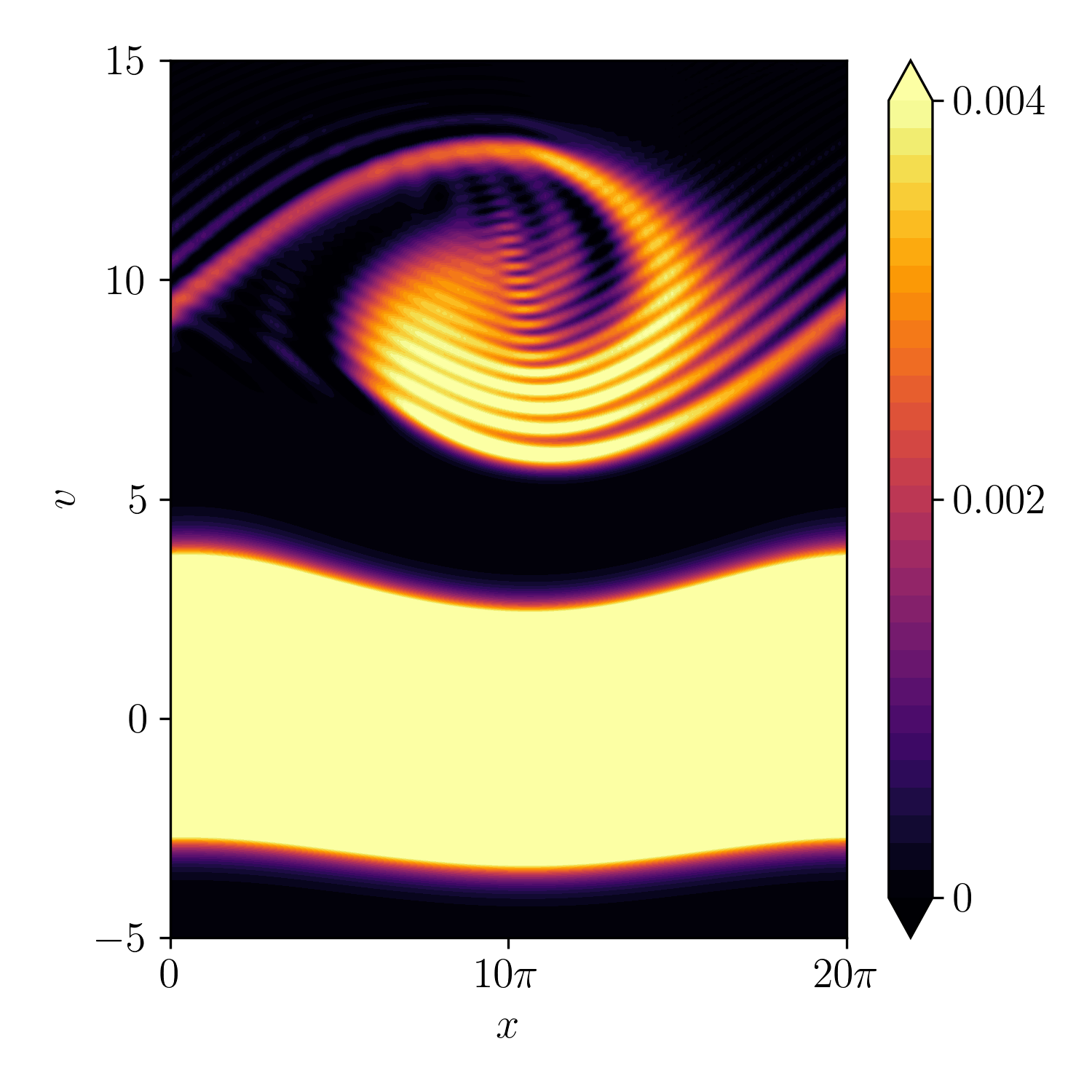}
    \end{subfigure}
    \caption{Bump-on-tail electron distribution function at $t= 90$ approximated via (a) Hermite, (b) Legendre, and (c) mixed method with 64 total DOFs in velocity. We also include (d) mixed method with 128 total DOFs in velocity and (e) reference solution obtained by a full finite differencing approach with $N_{v} = 60,000$.}
    \label{fig:bump_on_tail_phase_space}
\end{figure}

\begin{figure}
    \centering
    \includegraphics[width=0.7\linewidth]{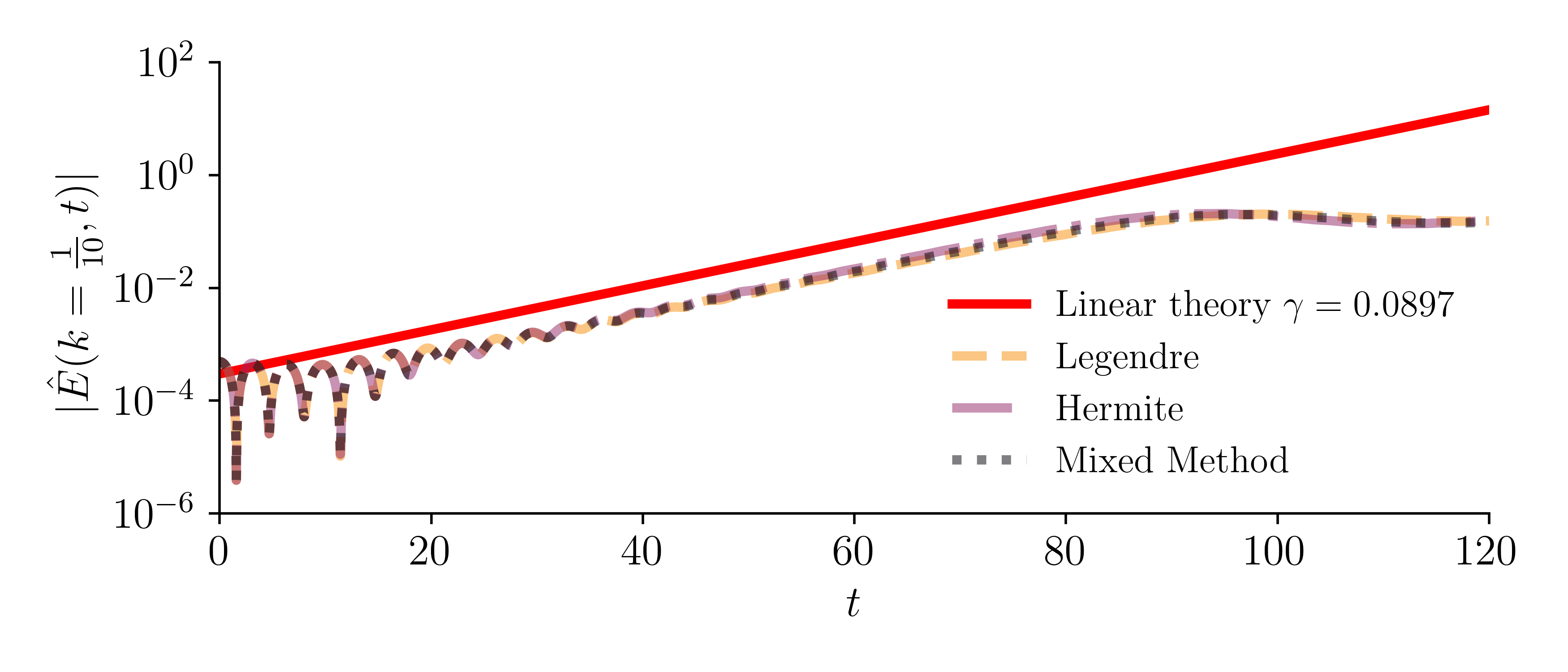}
    \caption{The bump-on-tail electric field first Fourier mode evolution in time for the three methods: Legendre with $N_{L} = 64$, Hermite $N_{H_{1}} = 16$ and $N_{H_{2}} = 48$, and mixed method with $N_{H} = 16$ and $N_{L} = 48$. The three methods are comparable and agree with the analytic growth rate $\gamma = 0.0897$ obtained from linear theory~\cite{gary_1993_theory, chapurin_2024_pop}.}
    \label{fig:E1-bump-on-tail}
\end{figure}

Figure~\ref{fig:bump_on_tail_phase_space} shows the electron distribution function at $t=90$ for the bump-on-tail instability, computed using the mixed method, the individual Hermite/Legendre methods, alongside the reference finite-difference solution with $N_v = 60{,}000$. For the Legendre simulation, we set $v_{a} = -5$ and $v_{b} = 15$.
Visually, the mixed method provides a closer match to the reference solution than either individual method (as quantified below), using the same total number of velocity DOFs, $N_v = 64$.  
This improved accuracy arises because the bulk of the distribution remains approximately Maxwellian, making the Hermite basis well-suited for capturing its dynamics, while the non-Maxwellian beam exhibits fine-scale filamentation and roll-up features that are more effectively resolved by the Legendre basis.  
By combining Hermite and Legendre expansions, the mixed method captures both the bulk and beam dynamics accurately, whereas the individual Hermite method struggles to resolve the beam, and the individual Legendre method fails to represent the bulk properly.  
Increasing the total velocity DOFs to $N_v = 128$ demonstrates convergence, with the mixed method accurately reproducing the detailed beam filamentation structures of the reference solution.
Moreover, examining the electric field first Fourier mode evolution in Figure~\ref{fig:E1-bump-on-tail} shows that all three methods produce comparable macroscopic dynamics and agree with the linear growth rate~\cite{gary_1993_theory, chapurin_2024_pop}.

\begin{figure}
    \centering
    \begin{subfigure}[b]{0.45\textwidth}
        \centering 
        \caption{Distribution function error}
        \includegraphics[width=\textwidth]{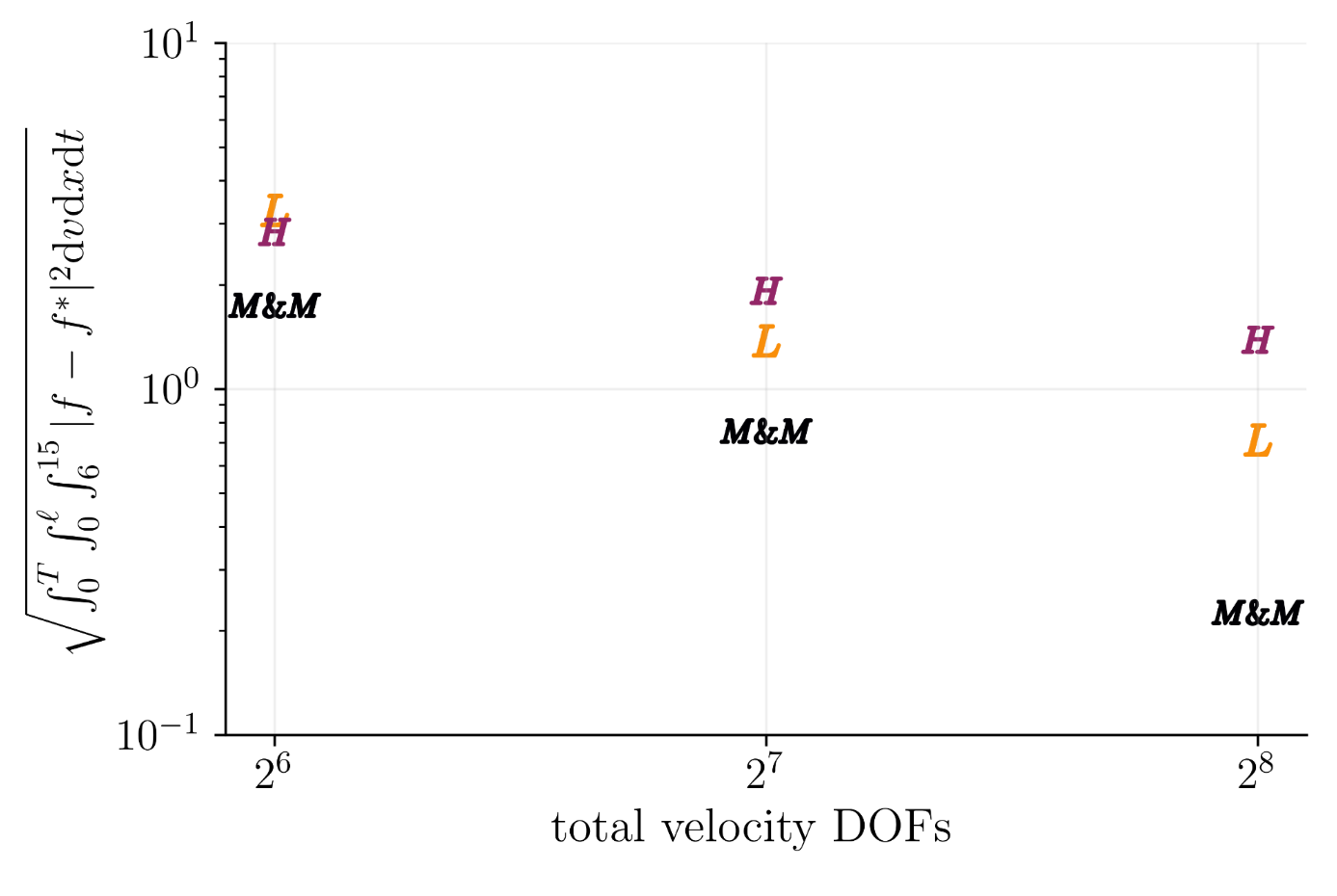}
    \end{subfigure}
    \hspace{-10pt}
    \begin{subfigure}[b]{0.45\textwidth}
        \centering 
        \caption{CPU runtime}
        \includegraphics[width=\textwidth]{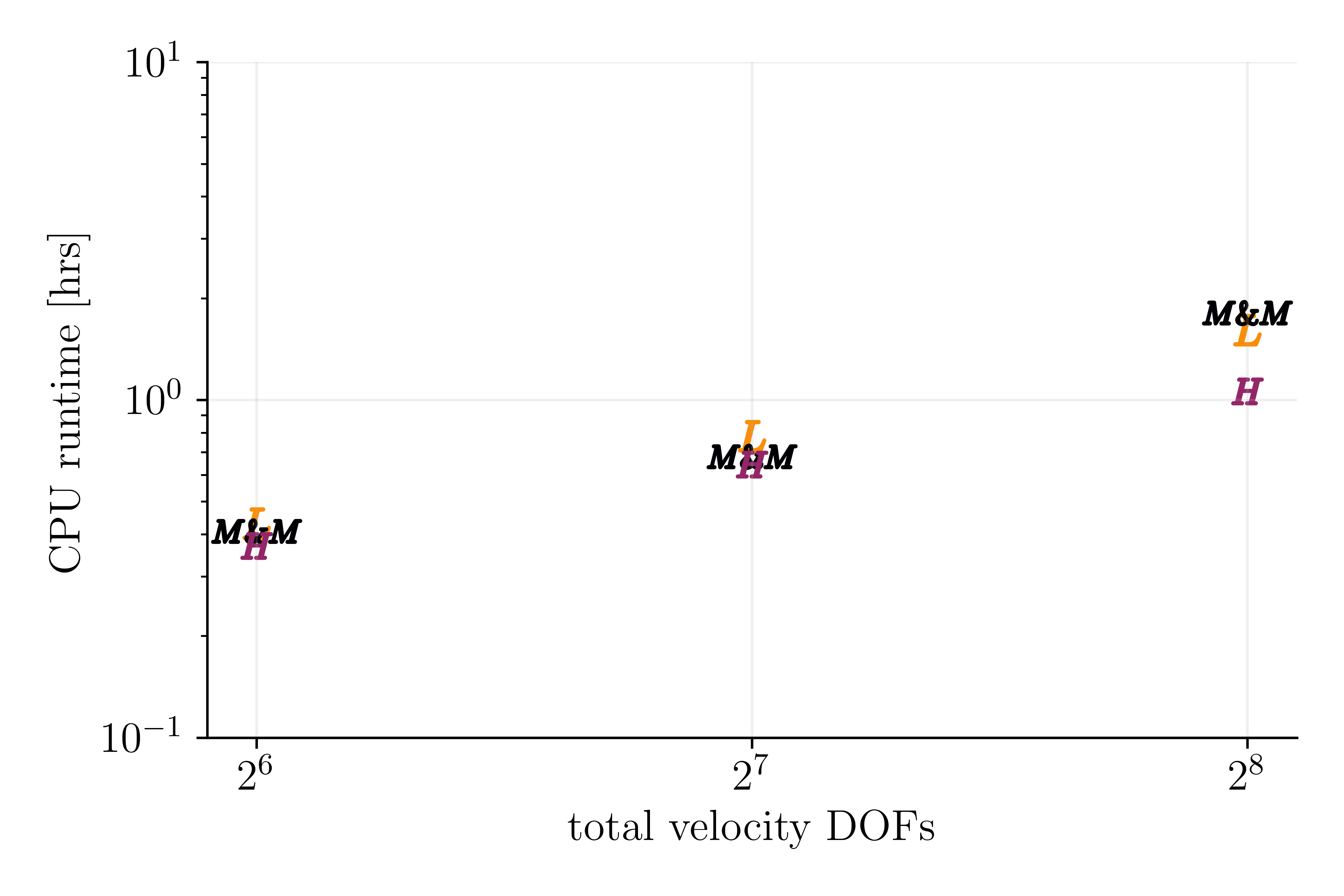}
    \end{subfigure}
    \caption{Same as Figure~\ref{fig:mixed_method_error_runtime_two_stream} for the bump-on-tail instability. Overall, the mixed method is more accurate than the individual Hermite and Legendre methods with comparable computational efforts.  }
    \label{fig:mixed_method_error_runtime_bump_on_tail}
\end{figure}

Lastly, similar to the two-stream instability problem, we can achieve improved accuracy with the mixed method in comparison to the individual Hermite and Legendre methods by restricting the Legendre description velocity bounds to a smaller domain (we use $v_{a} = 4$ and $v_{b} = 15$) as shown in Figure~\ref{fig:mixed_method_error_runtime_bump_on_tail}. 
We note that the mixed method appears to have a faster asymptotic convergence rate than the individual Hermite/Legendre methods, while the CPU runtime for all three methods is again comparable.

\section{Conclusions}\label{sec:conclusions}
In this paper, we have presented a mixed approach for the solution of the kinetic (Vlasov--Poisson) equations describing the dynamics of a plasma in the electrostatic limit. The method is based on splitting the distribution function in two parts (labeled as $f_0$ and $\delta f$), which are treated by two different discretizations in velocity space. 
The first is based on Hermite functions, which are suitable to capture near-Maxwellian behavior, while the second is based on Legendre polynomials, which can capture strong non-Maxwellian behavior.
A constraint in Eq.~\eqref{constraint} is applied to render the partition of phase space unique. 
Here we have chosen to employ simplicity, i.e. used the constraint that minimizes the explicit coupling between $f_0$ and $\delta f$: coupling terms arising from $f_0$ appear explicitly only in the equation for $\delta f$. 
(Note that, potentially, the last equation for the Hermite coefficient $C_{N_H-1}$ accounts for a term arising from $\delta f$ if the closure in Eq.~\eqref{closure_df} is used instead of the closure by truncation.)

The conservation laws of the mixed method have been derived in the semi-discrete and have been successfully verified with numerical experiments. When $N_H \ge 3$ and the velocity domain of the Legendre representation tends to infinity, the conservation laws are satisfied. When the Legendre velocity domain is finite (but symmetric), the conservation of total mass and energy (momentum) is satisfied if $N_H$ is even (odd). 
A slight modification of the algorithm has also been proposed to enforce the conservation laws exactly in all cases.
This can be accomplished by removing the coupling of the last Hermite function, $\psi_{N_H}$, with the first three Legendre terms.
For the examples considered in the paper this latter approach was very effective, as it enforced exact conservation laws while preserving the accuracy of the mixed method.

The mixed method has been tested with three benchmark problems: linear advection, two-stream instability and bump-on-tail instability. 
In all cases, the mixed method was compared against methods that use only one of the spectral representations (either Hermite or Legendre) and against a highly accurate reference solution obtained by central finite differences in velocity space.
The numerical experiments showed that the mixed method is not effective when the Legendre and Hermite representations span the full velocity domain, as in the linear advection test. 
In this case, the accuracy of the mixed method is dictated by the most accurate of the two representations and there is no gain in combining the two representations in a static manner.
In fact, for cases where strong non-Maxwellian features develop globally over time, starting from a Hermite-only representation and then switching completely to a Legendre-only representation would be more advantageous from the perspective of computational performance. 
The mixed method as formulated, on the other hand, becomes advantageous when strong non-Maxwellian behavior is local, confined to only a subset of velocity space, which would be very challenging to handle by the individual Hermite or Legendre representations alone. This aspect was exploited for the two-stream and bump-on-tail instability tests, where it was shown that the mixed method was more accurate for comparable computational performance.
As mentioned above, different mixed methods can be derived by imposing different partitions of phase space. Exploring such avenues will be important for the development of the optimal mixed method.


\section*{Acknowledgment}
O.I. was partially supported by the Los Alamos National Laboratory (LANL) Student Fellowship sponsored by the Center for Space and Earth Science (CSES). 
CSES is funded by LANL's Laboratory Directed Research and Development (LDRD) program under project number 20210528CR. 
O.I. was partially supported by the Strategic Enhancement of Excellence through Diversity Fellowship at the University of California, San Diego, in the Department of Mechanical and Aerospace Engineering. 
The LANL LDRD Program supported G.L.D. under project number 20250577ER and V.R. under project 20250500ER.
LANL is operated by Triad National Security, LLC, for the National Nuclear Security Administration of the US Department of Energy (Contract No. 89233218CNA000001).

\bibliography{references}

\end{document}